\documentclass[graybox, envcountchap]{svmult}

\usepackage{mathptmx}        % selects Times Roman as basic font
\usepackage{amsmath,amsfonts,amstext,amssymb}
\usepackage{color}
\usepackage{helvet}          % selects Helvetica as sans-serif font
\usepackage{courier}         % selects Courier as typewriter font
\usepackage{dirtree}
\usepackage{makeidx}        % allows index generation
\usepackage{graphicx}        % standard LaTeX graphics tool
\usepackage{subfig}

\usepackage{multicol}        % used for the two-column index
\usepackage[bottom]{footmisc}% places footnotes at page bottom

\usepackage{hyperref}        %for hyperlinks
\hypersetup{colorlinks=true,urlcolor=blue}

\usepackage[misc]{ifsym}

\makeindex             % used for the subject index
\usepackage{xspace}

\newcommand{\Mesh}{\texttt{Mesh}}
\newcommand{\MeshBlock}{\texttt{MeshBlock}}

\newcommand{\be}{\begin{eqnarray}}
\newcommand{\ee}{\end{eqnarray}}
\newcommand{\tGRAthena}{\texttt{GR-Athena++}}
\newcommand{\GRAthena}{\tGRAthena\xspace}

\newcommand{\tAthena}{\texttt{Athena++}}
\newcommand{\Athena}{\tAthena\xspace}

\newcommand{\tGRoAthena}{\texttt{(GR-)Athena++}}
\newcommand{\GRoAthena}{\tGRoAthena}

\newcommand{\tBAM}{\texttt{BAM}}
\newcommand{\BAM}{\tBAM\xspace}

\newcommand{\treprimand}{\texttt{RePrimAnd}}
\newcommand{\reprimand}{\treprimand\xspace}
\newcommand{\tPrimitiveSolver}{\texttt{PrimitiveSolver}}
\newcommand{\PrimitiveSolver}{\tPrimitiveSolver\xspace}

\def\d{{\rm d}}

\def\i{{\rm i}}

\def\gccm{{\rm g\,cm^{-3}}}

\def\GMc2{{\rm G M_{\odot} c^{-2}}}

\def\eps{\epsilon}
\def\eps{\epsilon}

\def\I{\mathcal{I}}

\def\l{\ell}

\def\kt2{\kappa^\text{T}_2}

\def\Mo{{\rm M_{\odot}}}
\def\kt2{\kappa^\text{T}_2}

\def\half{\frac{1}{2}}

\def\KO{\sigma_D}

\def\D{\mathrm{D}}
\def\pd{\partial{}}

\def\Ng{N_{\mathrm{g}}}

\def\2nd{2^\mathrm{nd}}
\def\4th{4^\mathrm{th}}
\def\6th{6^\mathrm{th}}
\def\8th{8^\mathrm{th}}
\def\sp{\delta x}
\def\defG{\widehat{\Gamma}}
\def\eg{{e.g.}}

\newcommand{\mc}[1]{\ensuremath{\mathcal{#1}}}

\def\z4c{$\mathrm{Z}4\mathrm{c}$}
\def\z4oc{$\mathrm{Z}4(\mathrm{c})$}
\def\z4{$\mathrm{Z}4$}
\def\ccz4{$\mathrm{CCZ}4$}

\begin{document}

\title{\tGRAthena{}: magnetohydrodynamical evolution with dynamical space-time}
\titlerunning{\tGRAthena{}: GRMHD with dynamical space-time}
\author{Boris Daszuta and William Cook}
\institute{Boris Daszuta (\Letter) \at Theoretisch-Physikalisches Institut, Friedrich-Schiller-Universit{\"a}t Jena, 07743, Jena, Germany, \email{boris.daszuta@uni-jena.de}
\and William Cook \at Theoretisch-Physikalisches Institut, Friedrich-Schiller-Universit{\"a}t Jena, 07743, Jena, Germany \email{william.cook@uni-jena.de}}
\maketitle

\abstract{%
We present a self-contained overview of \GRAthena{}, a general-relativistic magnetohydrodynamics (GRMHD) code, that incorporates treatment of dynamical space-time, based on the recent work of Daszuta+~\cite{Daszuta:2021ecf} and Cook+~\cite{Cook:2023bag}.
General aspects of the \Athena{} framework we build upon, such as oct-tree based, adaptive mesh refinement (AMR) and constrained transport, together with our modifications, incorporating the \z4c{} formulation of numerical relativity, judiciously coupled, enables GRMHD with dynamical space-times.
Initial verification testing of \GRAthena{} is performed through benchmark problems that involve isolated and binary neutron star space-times. This leads to stable and convergent results. Gravitational collapse of a rapidly rotating star through black hole formation is shown to be correctly handled.  In the case of non-rotating stars, magnetic field instabilities are demonstrated to be correctly captured with total relative violation of the divergence-free constraint remaining near machine precision.
The use of AMR is show-cased through investigation of the Kelvin-Helmholtz instability which is  resolved at the collisional interface in a merger of magnetised binary neutron stars.
The underlying task-based computational model enables \GRAthena{} to achieve strong scaling efficiencies above $80\%$ in excess of $10^5$ CPU cores and excellent weak scaling up to $\sim 5 \times 10^5$ CPU cores in a realistic production setup. \GRAthena{} thus provides a viable path towards robust simulation of GRMHD flows in strong and dynamical gravity with exascale high performance computational infrastructure.
}

\section{Introduction}
Embedding the techniques of numerical relativity (NR) within robust, highly-performant simulation infrastructure provides an indispensable tool for understanding the phenomenology of astrophysical inspiral and merger events. This is especially pertinent due to the arrival of multimessenger astronomy observations. Indeed recent, near-simultaneous detections of the gravitational wave signal GW170817, and associated electromagnetic counterpart GRB 170817A, alongside a kilonova, resulting from the merger of a Binary Neutron Star (BNS) system, have been made \cite{TheLIGOScientific:2017qsa,Goldstein:2017mmi,Savchenko:2017ffs}. These two events provide insight on the astrophysical origins of short Gamma Ray bursts (SGRB) as ensuing from BNS progenitors \cite{Monitor:2017mdv}.

The end products of BNS mergers have long been viewed as candidates for the launching of relativistic jets giving rise to SGRBs \cite{Blinnikov:1984a,Paczynski:1986px,Goodman:1986a,Eichler:1989ve,Narayan:1992iy} however the precise mechanism remains an open question. The presence of large magnetic fields in a post-merger remnant may play a role in jet formation \cite{Piran:2004ba,Kumar:2014upa,Ciolfi:2018tal}. Such fields may arise as the result of magnetohydrodynamic (MHD) instabilities during the late inspiral such as the Kelvin-Helmholtz instability (KHI) \cite{Rasio:1999a,Price:2006fi,Kiuchi:2015sga}, magnetic winding, and the Magneto-Rotational Instability (MRI) \cite{Balbus:1991a}. 
 The configuration of magnetic fields for a BNS system during the early inspiral is an open question. Already in the case of an isolated neutron star an initially poloidal field can give rise to instability \cite{Tayler:1957a,Tayler:1973a,Wright:1973a,Markey:1973a,Markey:1974a,Flowers:1977a}, and long-term simulations involving isolated stars suggest that higher resolution is required to understand reasonable initial configurations for the magnetic fields \cite{Kiuchi:2008ss,Ciolfi:2011xa,Ciolfi:2013dta,Lasky:2011un,Pili:2014npa,Pili:2017yxd,Sur:2021awe}, with thermal effects also playing a key role in the evolution of the field \cite{Pons:2019zyc}.

The distribution and nature of matter outflow from a BNS merger may also be influenced by the presence of magnetic fields \cite{Siegel:2014ita,Kiuchi:2014hja,Siegel:2017nub,Mosta:2020hlh,Curtis:2021guz,Combi:2022nhg,deHaas:2022ytm,Kiuchi:2022nin,Combi:2023yav}. These outflows determine the nature of the long lived electromagnetic signal following merger, such as AT2017gfo which accompanied GW170817 \cite{GBM:2017lvd,Coulter:2017wya,Soares-santos:2017lru,Arcavi:2017xiz}, known as the kilonova \cite{Li:1998bw,Kulkarni:2005jw,Metzger:2010sy}, as well as the r-process nucleosynthesis responsible for heavy element production occurring in ejecta \cite{Pian:2017gtc,Kasen:2017sxr}. 
It is clear that understanding the evolution of a BNS system from its late inspiral, through merger, to the post-merger is a formidable endeavour. A broad range of physical processes must be modeled consistently that include general relativity (GR), MHD, weak nuclear processes leading to neutrino emission and reabsorption, and the finite temperature behaviour of dense nuclear matter.

For the solution of the Einstein field equations, two commonly utilized classes of free evolution formulations of NR are: those based on moving puncture gauge conditions \cite{Brandt:1997tf,Campanelli:2005dd,Baker:2005vv}, such as BSSN \cite{Shibata:1995we,Baumgarte:1998te}, 
Z4c \cite{Bernuzzi:2009ex,Ruiz:2010qj,Weyhausen:2011cg,Hilditch:2012fp}, and CCZ4 formulations 
\cite{Alic:2011gg,Alic:2013xsa}; and the Generalised Harmonic Gauge approach
\cite{Pretorius:2005gq}. These may be coupled to MHD so as to provide a system of GRMHD evolution equations. In this context the ideal MHD limit of a non-resistive and non-viscous fluid is typically assumed. The aforementioned are written as a system of balance laws, which is key for shock capture and mass conservation \cite{Font:2007zz}. In order to preserve the magnetic field divergence-free condition a variety of approaches have been considered: Constrained Transport (CT) \cite{Evans:1988a,Ryu:1998ar,Londrillo:2003qi}; Flux-CT~\cite{Toth:2000}; vector potential formulation~\cite{Etienne:2011re}; and divergence cleaning ~\cite{Dedner:2002a}. Extensive effort has been invested into constructing codes for GRMHD simulation involving such techniques, a (non-exhaustive) list of examples include \cite{%
Baiotti:2004wn,%
Duez:2005sf,%
Shibata:2005gp,%
Anderson:2006ay,%
Giacomazzo:2007ti,%
Duez:2008rb,%
Liebling:2010bn,%
Foucart:2010eq,%
Thierfelder:2011yi,%
East:2011aa,%
Loffler:2011ay,%
Moesta:2013dna,%
Radice:2013xpa,%
Etienne:2015cea,%
Kidder:2016hev,%
Palenzuela:2018sly,%
Vigano:2018lrv,%
Cipolletta:2019geh,%
Cheong:2020kpv,%
Shankar:2022ful}. 
Aside from the complexity of the underlying physics modeled, a further important concern for NR based investigations is efficient resolution of solution features that develop over widely disparate scales that scale when simulations utilize high performance computing (HPC) infrastructure.

To this end we overview and demonstrate the recently developed GRMHD capabilities of our code \GRAthena{} \cite{Daszuta:2021ecf,Cook:2023bag}. We build upon a public version of \Athena{} \cite{Stone:2020} which is an astrophysical (radiation), GRMHD, {\tt c++} code for stationary space-times that features a task-based computational model built to exploit hybrid parallelism through dual use of message passing interface {\tt MPI} and threading via {\tt OpenMP}. Our efforts embed the \z4c{} formulation suitably coupled to GRMHD thereby removing the stationarity restriction. This allows \GRAthena{} to evolve magnetised BNS. To this end we leverage the extant \Athena{} infrastructure and methods such as CT. We inherit block-based, adaptive mesh refinement (AMR) capabilities which exploit (binary, quad, oct)-tree structure with synchronisation restricted to layers at block boundaries. This can be contrasted against other codes utilizing AMR with fully-overlapping, nested grids that rely on the Berger-Oliger~\cite{Berger:1984zza,Berger:1989a} time subcycling algorithm.

On account of this \GRAthena{} can efficiently resolve features developing during the BNS merger, such as small scale structures in the magnetic field instabilities, without requiring large scale refinement of coarser features. Performance scaling is demonstrated to $\mathcal{O}(10^5)$ CPU cores, showing efficient utilization of exascale HPC architecture.

\section{Code overview}
\GRAthena{} builds upon \Athena{} thus in order to specify nomenclature, provide
a self-contained description, and explain our extensions, we first briefly recount some details
of the framework (see also \cite{White:2015omx,%
felker2018fourthorderaccuratefinite,%
Stone:2020}).

In \Athena{}, overall details about the domain $\Omega$ over which a problem is
formulated are abstracted from the salient physics and contained
within a class called the \Mesh{}.
Within the \Mesh{} an overall representation of
the domain as a logical $n$-rectangle is stored, together with details of
coordinatization type (Cartesian or more generally curvilinear), number of points along each
dimension for the coarsest sampling $N_M=(N_{M_1},\,\cdots,\,N_{M_d})$, and physical boundary
conditions on
$\partial \Omega$. Next, we would like to decompose the domain into smaller blocks, described by so-called {\tt MeshBlock} objects. In order to partition the domain
we first fix a choice $N_B=(N_{B_1},\,\cdots,\,N_{B_d})$ where each element of $N_B$ must divide
each element of $N_M$ component-wise. Then $\Omega$ is domain-decomposed through
rectilinear sub-division into a family of $n$-rectangles satisfying
$\Omega = \sqcup_{i\in Z} \Omega_i$, where $Z$ is the set of \MeshBlock{} indices, corresponding to the ordering described in \S \ref{ssec:tree_structure}.
Nearest-neighbor elements are constrained to only differ by a single sub-division at most.
The \MeshBlock{} class stores properties of an element $\Omega_i$ of the
sub-division. In particular the number of points in the sampling of
$\Omega_i$ is controlled through the choice of $N_B$. For purposes of communication of data between
nearest neighbor \MeshBlock{} objects the sampling
over $\Omega_i$ is extended by a thin layer of so-called ``ghost nodes'' in each direction.
Furthermore the local values (with respect to the chosen, extended sampling
on $\Omega_i$) of any discretized, dependent field variables of interest are stored
within the \MeshBlock{}.

In both uniform grid $(\forall i\in Z)$ $\mathrm{vol}(\Omega_i) = C$ and refined meshes
$(\exists i,\,j\in Z)$ $\mathrm{vol}(\Omega_i) \neq \mathrm{vol}(\Omega_j)$ it is
crucial to arrange
inter-\MeshBlock{} communication efficiently -- to this end the relationships
between differing \MeshBlock{} objects are arranged in a tree data structure,
to which we now turn.

\subsection{Tree structure of the \Mesh{}}\label{ssec:tree_structure}
For the sake of exposition
here and convenience in later sections we now particularize to a
Cartesian coordinatization though we emphasize that the general
picture (and our implementation) of the discussions here and in \S\ref{ssec:field_discretization}
carry over to the curvilinear context with only minor modification.

\GRoAthena{} stores the logical relationship between the \MeshBlock{} objects (i.e. $\Omega_i$)
involved in description of a domain $\Omega$ within a tree data structure. A binary-tree,
quad-tree
or oct-tree is utilized when $d:=\dim(\Omega)=1,\,2,\,3$ respectively. The relevant
tree is then
constructed by first selecting the minimum $N$ such that $2^N$ exceeds the largest number of
$\Omega_i$ along any dimension. The root of the tree is assigned a logical level of zero and the
tree terminates at level $N$ with every \MeshBlock{} assigned to an appropriate leaf, though
some leaves and nodes of the tree may remain empty. Data locality is enhanced, as references to
\MeshBlock{} objects are stored such that a post-order,
depth-first traversal of the tree follows Morton order (also termed Z-order)
\cite{morton1966computer}.
This order can be used to encode multi-dimensional coordinates into a linear index
parametrizing a Z-shaped, space-filling curve where small changes in the parameter imply spatial
coordinates that are close in a suitable sense \cite{burstedde2019numberfaceconnectedcomponents}.

\begin{figure}[b]
  \sidecaption
	\centering
  \includegraphics[width=6cm]{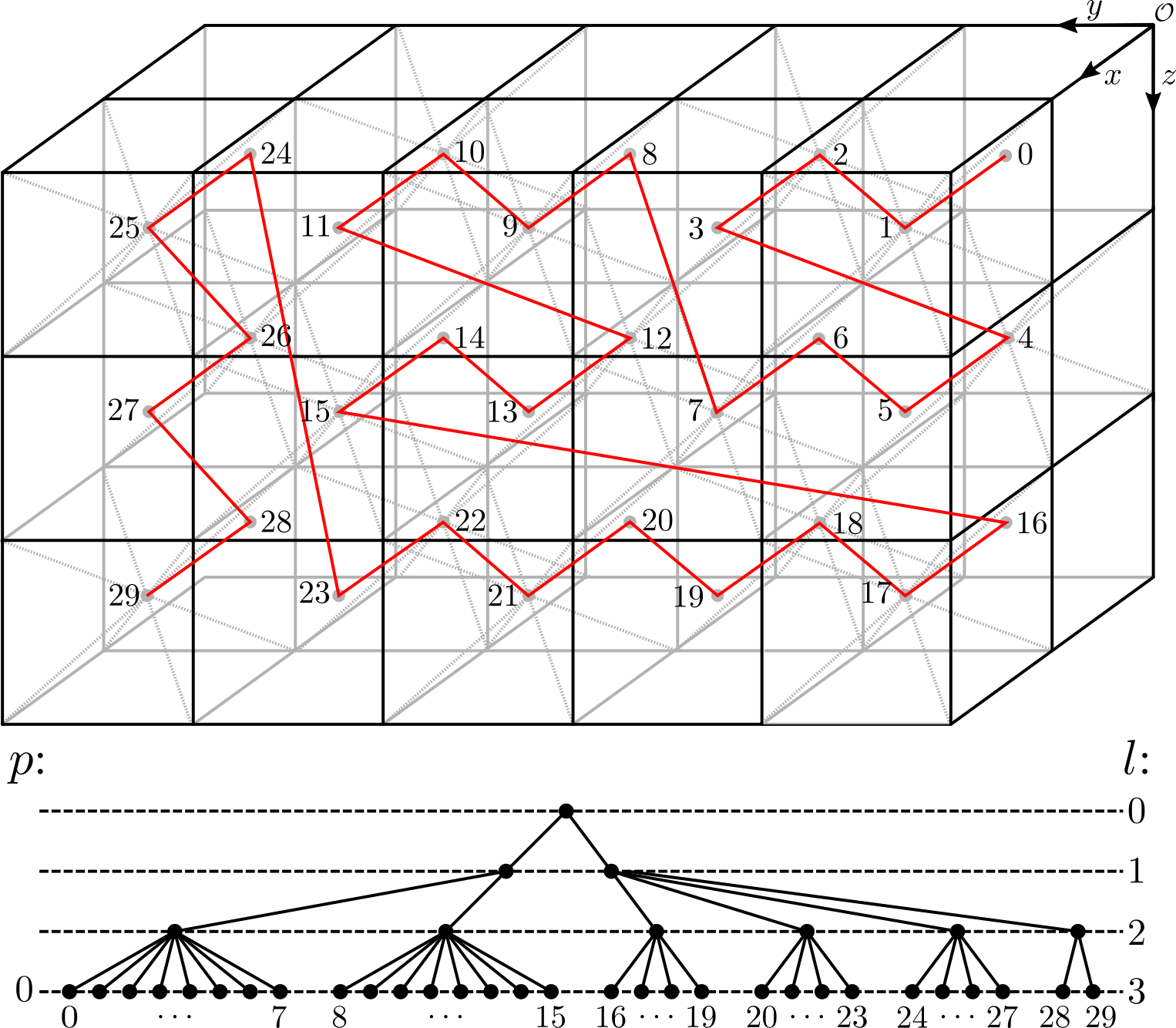}
  \caption{%
  Example of \Mesh{} partitioned uniformly by \MeshBlock{} objects indexed via Z-order and
  traced in red through constituent geometric centroids.
  The logical relationship
  between $\Omega_i$ is stored in an oct-tree. Empty leaves are suppressed though each populated
  node above logical level three has eight children.
  Notice that physical level $p$ and logical level $l$ are distinct.
  (Figure adapted from \cite{Daszuta:2021ecf}).
  }
\label{fig:oct-tree_partitioned}
\end{figure}

As an example we consider a three-dimensional \Mesh{} described by
$(N_x,\,N_y,\,N_z)=(2,\,5,\,3)$ \MeshBlock{}
objects in each direction at fixed physical level in Fig.\ref{fig:oct-tree_partitioned}.

\subsubsection{{\tt Mesh} refinement}\label{sssec:mesh_refinement}
In order to resolve solution features over widely varying spatial (temporal) scales \GRoAthena{} implements block-based adaptive mesh refinement (AMR) \cite{stout1997adaptiveblockshigh}. This naturally fits into the domain-decomposition structure previously described as each physical position on a computational domain is covered by one and only one level. Salient field data need only be synchronized at block ({\tt MeshBlock}) boundaries. The logical arrangement into a (binary,quad,oct)-tree structure has the added benefit of improving computational efficiency through preservation of data locality in memory. Similarly the underlying task-based parallelism (see \S\ref{ssec:task_list}) as the computational model greatly facilitates the overlap of communication and computation.

Consider now a \Mesh{} with refinement. The number of
physical refinement levels added to a uniform level, domain-decomposed $\Omega$ is
controlled by the parameter $N_L$.
By convention $N_L$ starts at zero. Subject to satisfaction of
problem-dependent, user-prescribed, refinement criteria, there may exist physical levels at $0,\,\cdots,\,N_L$.
When a given \MeshBlock{} is refined (coarsened) $2^d$ \MeshBlock{} objects are constructed
(destroyed). This is constrained to satisfy a $2:1$ refinement ratio where
nearest-neighbor \MeshBlock{} objects can differ by at most one physical level. Function data at a fixed physical level is transferred one level finer through use of a prolongation operator $\mathcal{P}$; dually, function data may be coarsened by one physical level through restriction $\mathcal{R}$. The details of the $\mathcal{P}$ and $\mathcal{R}$ operators depend on the underlying field treated, together with the selection of discretization.

\begin{figure}[b]
  \sidecaption
	\centering
  \includegraphics[width=6cm]{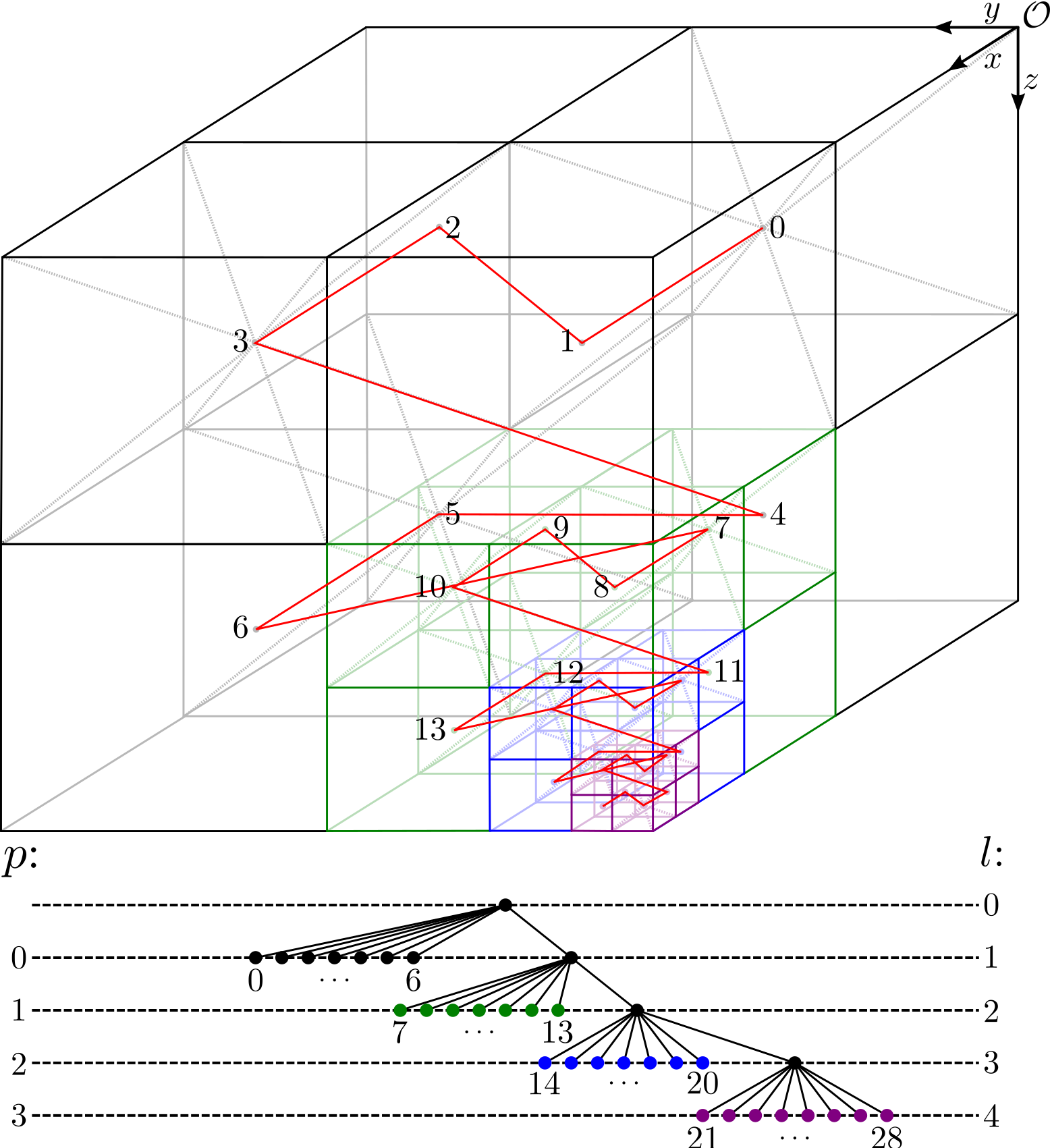}
  \caption{%
  Example of \Mesh{} partitioned and refined by \MeshBlock{} objects indexed
    via Z-order and
  traced in red through constituent geometric centroids. The logical relationship
  between ${}^p\Omega_i$ and neighbors
  is stored in an oct-tree. There are no unpopulated leaves.
  Notice that physical level $p$ and logical level $l$ are distinct; coloring corresponds to
  physical level: $p=0$ in black, $p=1$ in dark green, $p=2$ in blue, and $p=3$ in purple.
  (Figure adapted from \cite{Daszuta:2021ecf}).
    }
  \label{fig:oct-tree_refined}
\end{figure}

A concrete example of a potential overall structure is provided in Fig.\ref{fig:oct-tree_refined} where we consider a non-periodic $\Omega$ described by
$N_x=N_y=N_z=2$ \MeshBlock{} objects with $N_L=3$ selected with refinement introduced
at the corner $x_{\max}$, $z_{\max}$. If periodicity conditions are imposed on $\partial \Omega$ then additional refinement may be
required for boundary intersecting \MeshBlock{} objects so as to maintain the aforementioned
inter-\MeshBlock{} $2:1$ refinement ratio.

Two main strategies for control over refinement are offered. One may pose that a fixed, region contained within the {\tt Mesh} is at a desired level $N_L$. Alternatively, on individual {\tt MeshBlock} objects a refinement condition may be imposed. This latter is quite flexible and allows for e.g.~time-dependent testing as to whether dynamical fields satisfy some desired criteria. In particular, in the binary-black-hole context, recent work with \GRAthena{} \cite{rashti2023adaptivemeshrefinement} has compared ``box-in-box'', ``sphere-in-sphere'', and truncation error estimation as ingredients for the aforementioned condition and the effect on gravitational wave mismatch.

\subsection{Field discretization}\label{ssec:field_discretization}
A variety of discretizations of dynamical field components are required to facilitate the evolution of the GRMHD system embedded (\S\ref{ssec:dynamical_system}) within \GRAthena{}. Natively \Athena{} supports cell-centered (CC), face-centered (FC), and edge-centered (EC) grids. Thus volume averaged, conserved (primitive) quantities of the Euler equations are taken on CC and suitable mapping to FC (during calculation of e.g.~fluxes) leads to a shock-capturing, conservative, Godunov-style finite volume method. When magnetic (electric) fields are present then EC is employed so as to enable use of the Contrained Transport algorithm (see \S\ref{ssec:constrained_transport}) which preserves the divergence free condition on magnetic fields.

To make the connection to a {\tt MeshBlock} object concrete, we can first consider the one-dimensional domain $\Omega_I:=[a,\,b]$. Grids for uniformly sampled CC and vertex-centers (VC) respectively are:
\begin{gather}\label{eq:grid_sampling}
  \begin{aligned}
    \mathcal{G}_{\mathrm{CC}} :=& \Big\{a + \Big(n + \frac{1}{2}\Big) \frac{b-a}{N_B}\,\Big|\,
    n \in \{0,\,\dots N_B-1\}\Big\},\\
    \mathcal{G}_{\mathrm{VC}} :=& \Big\{a + n \frac{b-a}{N_B}\,\Big|\,
      n \in \{0,\,\dots N_B\}\Big\};
  \end{aligned}
\end{gather}
where $\sp{}:=(b-a)/N_B$ is the grid spacing and $N_B$ the sampling parameter. To facilitate communication between {\tt MeshBlock} objects a thin layer of $\Ng{}$ ``ghost nodes'' (also with spacing $\sp$) is extended in the direction of each nearest-neighbor {\tt MeshBlock}. Products of combinations of such grids allow for straight-forward extension to $2^\mathrm{d}$ and $3^{\mathrm{d}}$, together with construction of e.g.~FC sampling.

For the evolution of purely geometric quantities (induced spatial metric, extrinsic curvature, etc) \GRAthena{} extends the framework by incorporating infrastructure that allows for point-wise description of fields on VC. Our primary motivation for this is one of computational efficiency which may be understood through the following. Consider splitting a fixed $\mathcal{G}_{\mathrm{VC}}$ exactly in half into two new grids each with the same sampling parameter as the parent. This doubles the resolution. We also find that the coarsely spaced parent nodes exactly coincide with interspersed nodes of the children. This behaviour occurs at refinement boundaries (in ghost-layers) for VC sampling due to the $2:1$ ratio nearest-neighbor {\tt MeshBlock} objects must obey (see \S\ref{sssec:mesh_refinement}). For refinement based on polynomial interpolation this reduces the restriction operation for VC to an injection (copy), and, simplifies prolongation. For a full discussion of our high-order $\mathcal{R}$ and $\mathcal{P}$ for VC see \cite{Daszuta:2021ecf}.

\subsubsection{Geodesic spheres}
\label{ssec:geodesic_spheres}

In \GRAthena{}, during simulations, evaluation of diagnostic quantities that are naturally defined as integrals over spherical surfaces is often required. As a particular example we extract gravitational radiation (\S\ref{sssec:wave_extraction}) through evaluation of numerical quadrature over triangulated geodesic spheres. Using a geodesic grid ensures more even tiling of the sphere compared to the uniform latitude-longitude grid of similar resolution.

\begin{figure}[b]
  \sidecaption
  \centering
  \includegraphics[width=4.5cm]{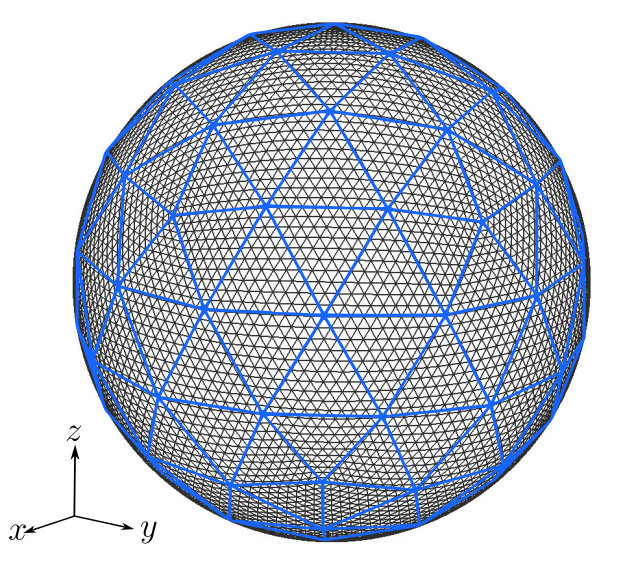}
  \caption{Structure of the geodesic grid used by \GRAthena{}. Depicted in light-blue is a grid in consisting of 92 vertices ($n_Q=3$). In black we show a grid of 9002 vertices ($n_Q=30$); a typical sampling associated with gravitational wave extraction in production simulations.
  }
  \label{fig:geogrid}
\end{figure}

A geodesic sphere of radius $R$ (denoted $Q_R$) may be viewed as the boundary of a convex polyhedron embedded in $\mathbb{R}^3$ with triangular faces, i.e., a simplicial $2$-sphere that is homeomorphic to the $2$-sphere of radius $R$.
To construct the geodesic grid we start from a regular icosahedron with 12 vertices and 20 plane equilateral triangular faces, embedded in a unit sphere.
This is refined using the so called ``non-recursive'' approach \cite{Wang:2011}. This leads to a polyhedron with $10n_Q^2+2$ vertices where the refinement parameter $n_Q$ controls the so-called grid level. In Fig.\ref{fig:geogrid} we show an example of two such grids.

Note that field data over the {\tt Mesh} (i.e.~local to {\MeshBlock} objects) is treated as independent from a geodesic sphere in our implementation, with transfer achieved via standard Lagrange interpolation.

For evaluation of numerical quadratures we associate to each grid point of a geodesic sphere a solid angle by connecting the circumcenters of any pair of triangular faces that share a common edge.
The solid angles subtended by the cells at the center of the sphere are used as weighting coefficients when computing the averages. The logical connection between the neighboring cells is implemented as described in
\cite{Randall:2002}.

\section{GRMHD}
\subsection{Dynamical system}\label{ssec:dynamical_system}
Our treatment of GRMHD within \GRAthena{} involves the \z4c{} system (\S\ref{sssec:z4c_system}) coupled to the equations of ideal MHD through matter sources (\S\ref{sssec:mhd_system}). The setting of the formulation utilizes the standard geometric view (and definitions) of ADM decomposition \cite{Arnowitt:1959ah} whereby a space-time manifold endowed with metric $(\mathcal{M},\,g)$ is viewed as foliated by a family of non-intersecting, spatial hypersurfaces with members $\Sigma_t$. Fixing notation, recall that one introduces a future-directed $t{}^a$ satisfying $t{}^a\nabla{}_a[t]=1$ and
considers $t{}^a=\alpha n{}^a+\beta{}^a$ where $n^a$ is a future-directed, time-like,
unit normal $n^a$ to each member of the foliation $\Sigma{}_t$, $\alpha$ is the
lapse and $\beta{}^a$ the shift. Projections of ambient fields over $\mathcal{M}$ to (various products of) the tangent and normal bundles of $\Sigma$ leads to evolution, and constraint equations. Here we summarise the resulting equations\footnote{Ambient quantities take indices $a,\,b,\,\ldots$ and $\mu,\,\nu,\,\ldots$ whereas spatial quantities take $i,\,j,\,\ldots$.} as we have implemented them.

\subsubsection{\z4c{} system}\label{sssec:z4c_system}
In brief, to stabilize the time-evolution problem, the \z4{} formulation \cite{Bona:2003fj} directly augments the Einstein field equations via suitable introduction of an auxiliary, dynamical vector field $Z^a$. This allows for derivation of evolution equations for the induced metric $\gamma{}_{ij}$ on $\Sigma$, the extrinsic curvature $K{}_{ij}:=-\frac{1}{2}\pounds{}_{\mathbf{n}}[\gamma{}_{ij}]$ (with $\pounds{}_{\mathbf{n}}[\cdot]$ the Lie derivative along $n{}^a$), and the projected auxiliary quantities $\Theta:=-n{}_a Z{}^a$ and $\check{Z}_i:=\perp^a_i Z{}_a$ (with $\perp^a_b:=g{}^a{}_b+n{}^a n{}_b$).

At its core \z4c{} \cite{Bernuzzi:2009ex,Hilditch:2012fp} is a conformal reformulation featuring constraint damping \cite{%
Gundlach:2005eh,%
Weyhausen:2011cg}. A further, spatial conformal degree of freedom is factored out:
\begin{align}\label{eq:confxform}
  \tilde{\gamma}{}_{ij} :=& \psi{}^{-4}\gamma{}_{ij}, &
  \tilde{A}{}_{ij}:=& \psi^{-4}\Big(K{}_{ij} - \frac{1}{3}K\gamma{}_{ij}\Big);
\end{align}
with the trace defined as $K:=K{}_{ij}\gamma{}^{ij}$, and conformal factor $\psi:=(\gamma/f)^{1/12}$ where $\gamma$ and $f$ are determinants of $\gamma{}_{ij}$ and some spatial reference metric $f{}_{ij}$ respectively.
It is assumed that $f{}_{ij}$ is flat and in Cartesian
coordinates which immediately yields the \emph{algebraic constraints}\footnote{We enforce $\mathcal{C}_A=0$ during numerical evolution to ensure consistency \cite{Cao:2011fu}.}:
\begin{align}\label{eq:algConstr}
  \mathcal{C}_A:=
  \big(
    \ln(\tilde{\gamma}),\,\tilde{\gamma}{}^{ij} \tilde{A}{}_{ij}
  \big)=0.
\end{align}
Additionally we introduce the definitions:
\begin{align}\label{eq:defn_transforms}
  &\chi:=\gamma{}^{-1/3}, &
  &\hat{K}:= K-2\Theta; &
  &\tilde{\Gamma}{}^i := 2\tilde{\gamma}{}^{ij} \check{Z}{}_j
  + \tilde{\gamma}{}^{ij}\tilde{\gamma}{}^{kl} \pd{}_l[\tilde{\gamma}{}_{jk}],&
  &\defG{}^i := \tilde{\gamma}{}^{jk}\tilde{\Gamma}{}^i{}_{jk}.
\end{align}
The dynamical variables
$\big(%
  \chi,\,\tilde{\gamma}{}_{ij},\,\hat{K},\tilde{A}{}_{ij},\,\Theta,\,\tilde{\Gamma}{}^i%
\big)$ obey the \emph{evolution equations}:
\begin{align}
  \pd{}_t[\chi] =& \frac{2}{3}\chi
  \left(
    \alpha(\hat{K} + 2\Theta) - \partial_i[\beta{}^i]
  \right) + \beta{}^i\partial_i[\chi],
  \label{eq:evo_chi}
\end{align}
\begin{equation}
  \pd{}_t[\tilde{\gamma}{}_{ij}] =
  -2\alpha\tilde{A}{}_{ij} + \beta{}^k\partial{}_k[\tilde{\gamma}{}_{ij}]
  -\frac{2}{3}\tilde{\gamma}{}_{ij}\partial{}_k[\beta{}^k]
  +2\tilde{\gamma}{}_{k(i}\partial{}_{j)}[\beta{}^k].
  \label{eq:evo_conf_metr}
\end{equation}
\begin{align}\nonumber
  \pd{}_t[\hat{K}] =&
  -\D{}^i[\D{}_i[\alpha]]
  + \alpha\left[
    \tilde{A}{}_{ij}\tilde{A}{}^{ij} + \frac{1}{3}(\hat{K}+2\Theta)^2
  \right]\\
  &
  + \beta{}^i \pd{}_i[\hat{K}]
  +\alpha\kappa{}_1(1-\kappa{}_2)\Theta+4\pi\alpha[{}^{~A}S+{}^{~A}\rho],
  \label{eq:evo_Khat}
\end{align}
\begin{align}\nonumber
  \pd{}_t[\tilde{A}{}_{ij}] &=
  \chi[-\D{}_i[\D{}_j[\alpha]] + \alpha(R{}_{ij}
  -8\pi {}^{~A}S{}_{ij})]^{\mathrm{tf}}
  +\alpha[
    (\hat{K} + 2\Theta)\tilde{A}{}_{ij}
    -2\tilde{A}{}^k{}_i\tilde{A}{}_{kj}
  ]
  \\
  &\hphantom{=}
  + \beta{}^k \pd{}_k[\tilde{A}_{ij}]
  + 2\tilde{A}{}_{k(i} \pd{}_{j)}[\beta{}^k]
  - \frac{2}{3}\tilde{A}{}_{ij}\pd{}_k[\beta{}^k],
  \label{eq:evo_conf_Atil}
\end{align}
\begin{align}
  \pd{}_t[\Theta] =& \frac{\alpha}{2}
  \left[
    \tilde{\mathcal{H}} -2\kappa{}_1(2+\kappa{}_2) \Theta
  \right] + \beta{}^i \pd{}_i[\Theta],
  \label{eq:evo_Theta}
\end{align}
\begin{align}
  \nonumber
  \pd{}_t[\tilde{\Gamma}{}^i] =&
  -2\tilde{A}{}^{ij} \pd{}_j[\alpha]
  +2\alpha
  \Big[
    \tilde{\Gamma}{}^i{}_{jk} \tilde{A}{}^{jk}
    -\frac{3}{2}\tilde{A}{}^{ij} \pd{}_j[\ln(\chi)]
    -\kappa{}_1(\tilde{\Gamma}{}^i-\defG{}^i)\\
    \nonumber
  &
  -\frac{1}{3}\tilde{\gamma}{}^{ij}\pd{}_j[2\hat{K}+\Theta]
  -8\pi\tilde{\gamma}{}^{ij}{}^{~A}S{}_j
  \Big]
  +\tilde{\gamma}{}^{jk}\pd{}_k[\pd{}_j[\beta{}^i]]
  +\frac{1}{3}\tilde{\gamma}{}^{ij}\pd{}_j[\pd{}_k[\beta{}^k]]\\
  &
  +\beta{}^j \pd{}_j[\tilde{\Gamma}{}^i]
  -\defG{}^j \pd{}_j[\beta{}^i]
  +\frac{2}{3}\defG{}^i \pd{}_j[\beta{}^j];
  \label{eq:evo_conf_GamTil}
\end{align}
where in Eq.\eqref{eq:evo_conf_Atil} the trace-free (tf) operation is computed with respect to
$\gamma{}_{ij}$ and $\tilde{\mathcal{H}}$ is defined in Eq.\eqref{eq:constr_ham}. The scalars $\kappa_1$ and $\kappa_2$ are freely chosen, damping parameters. Definitions of matter fields are standard and based on projections of the decomposed space-time,
energy-momentum-stress tensor $T{}_{ab} = {}^A\rho n{}_a n{}_b + 2 {}^AS{}_{(a} n{}_{b)} + {}^AS{}_{ab}$
in terms of the energy density ${}^A\rho:=T{}_{ab} n{}^a n{}^b$, momentum ${}^AS{}_i=-T{}_{bc} n{}^b \perp^c_i$, and spatial stress ${}^AS{}_{ij}:= T{}_{cd} \perp^c_i \perp^d_j$.
The intrinsic curvature $R{}_{ij}$ appearing in Eq.\eqref{eq:evo_conf_Atil} is decomposed according to $R{}_{ij} = \tilde{R}^\chi{}_{ij} + \tilde{R}{}_{ij}$
where in terms of the connection $\tilde{\mathrm{D}}{}_i$ compatible
with $\tilde{\gamma}{}_{jk}$:
\begin{equation*}
  \tilde{R}{}^\chi{}_{ij} =
  \frac{1}{2\chi}\left[
    \tilde{\D}{}_i[\tilde{\D}{}_j[\chi]]
    +\tilde{\gamma}{}_{ij}\tilde{\D}{}^l[\tilde{\D}{}_l[\chi]]
    -\frac{1}{2\chi}\tilde{\D}{}_i[\chi]\tilde{\D}{}_j[\chi]
  \right]
  -\frac{3}{4\chi^2} \tilde{\D}{}^l[\chi]\tilde{\D}{}_l[\chi] \tilde{\gamma}{}_{ij},
  \end{equation*}
and:
\begin{equation*}
  \tilde{R}{}_{ij} =
  -\frac{1}{2}\tilde{\gamma}{}^{lm}
  \pd{}_l[\pd{}_m[\tilde{\gamma}{}_{ij}]]
  +\tilde{\gamma}{}_{k(i}
  \pd{}_{j)}[\tilde{\Gamma}{}^k]
  +\defG{}^k\tilde{\Gamma}{}_{(ij)k}
  +\tilde{\gamma}{}^{lm}
  (
    2\tilde{\Gamma}{}^k{}_{l(i}\tilde{\Gamma}{}_{j)km}
    +\tilde{\Gamma}{}^k{}_{im}\tilde{\Gamma}{}_{klj}
  ).
\end{equation*}

The \emph{dynamical constraints} in terms of transformed variables
$(\tilde{\mathcal{H}},\,\tilde{\mathcal{M}}{}_i,\,\Theta,\,\check{Z}{}^i)$
may be monitored\footnote{Alternatively these may be assessed mapping back from the conformal variables.} to assess the quality of a numerical calculation:
\begin{equation}
  \tilde{\mathcal{H}} := R - \tilde{A}{}_{ij} \tilde{A}{}^{ij}
  +\frac{2}{3}\big(
  \hat{K}+2\Theta
	\big)^2 - 16\pi {}^{~A}\rho = 0,
  \label{eq:constr_ham}
\end{equation}
\begin{equation}
  \tilde{\mathcal{M}}{}_j :=
  \tilde{\mathrm{D}}_i[\tilde{A}{}^{i}{}_j]
  - \frac{3}{2}\tilde{A}{}^i{}_j\partial{}_i[\ln(\chi)]
	-\frac{2}{3}\pd{}_j[\hat{K}+2\Theta] - 8\pi {}^{~A}S_j =0,
\end{equation}
\begin{align}
  \Theta=&0, &
  \check{Z}{}^i=& \tilde{\Gamma}{}^i - \defG{}^i =0.
\end{align}

In order to close the \z4c{} system we must supplement it with gauge conditions. These are conditions on $(\alpha,\,\beta^i)$ that in \GRAthena{} are based on Bona-M\'asso lapse
\cite{Bona:1994b} and the gamma-driver shift \cite{Alcubierre:2002kk}:
\begin{align}\label{eq:gaugeBGDRV}
  \partial_t[\alpha] &= -\mu{}_L \alpha^2 \hat{K} +
    \beta{}^i \partial_i[\alpha],
  &
  \partial_t[\beta{}^i] &= \mu{}_S \alpha^2 \tilde{\Gamma}{}^i
    -\eta \beta{}^i + \beta{}^j \partial{}_j[\beta{}^i].
\end{align}
In specification of Eq.\eqref{eq:gaugeBGDRV} we employ the $1+\log$ lapse variant $\mu{}_L=2/\alpha$ together with $\mu{}_S=1/\alpha{}^2$. The $\eta$ term serves as a free damping parameter.

\subsubsection{MHD and coupling}\label{sssec:mhd_system}

In order to couple to the \z4c{} system of \S\ref{sssec:z4c_system} consider the energy-momentum-stress (EMS) tensor factored into perfect fluid and electromagnetic parts:
\begin{equation}\label{eq:emsMHD}
  T{}^{\mu\nu} = \Big[
    \rho h u{}^\mu u{}^\nu + p g{}^{\mu\nu}
  \Big] +
  \Big(
    b{}^2 u{}^\mu u{}^\nu + \frac{1}{2}b^2 g{}^{\mu\nu}
    -b{}^\mu b{}^\nu
  \Big).
\end{equation}
Focusing on the perfect fluid sector (first bracketed term) we have fluid quantities $u{}^\mu$, $\rho$, $p$, $h:=1+\epsilon+p/\rho$, and $\epsilon$, which are respectively a time-like velocity 4-vector, rest mass density, pressure, relativistic specific enthalpy, and specific internal energy. Write $u{}^\mu =(u{}^0, u{}^i)$ and define spatial components of the fluid $4$-velocity by forming the projection to $\Sigma_t$ as:
\begin{equation}
  \tilde u{}^i = \bot{}^i_\mu u{}^\mu = u{}^i + \frac{W \beta{}^i}{\alpha},
\end{equation}
where $W:=-n{}_\mu u{}^\mu$ is the Lorentz factor between Eulerian and comoving fluid observers. This leads to the fluid $3$-velocity $v{}^i:=\tilde{u}{}^i /W$ and $\tilde{v}{}^i:=v{}^i-\beta{}^i/\alpha$. The second bracketed term of Eq.\eqref{eq:emsMHD} carries the electromagnetic contribution. The terms $b{}^\mu$ and $b^2$ appearing there describe the magnetic field and arise as projections of the dual Faraday tensor along $u{}^\mu$. Recall that with respect to Eulerian observers the magnetic $\mc{B}^\mu$ and electric field $\mc{E}^\mu$ can be used to express the Faraday tensor as:
\begin{equation*}
  F{}^{\mu\nu} =
  n{}^\mu \mc{E}{}^\nu - \mc{E}{}^\mu n{}^\nu -
  \eps{}^{\mu\nu\rho\sigma} n{}_\rho \mc{B}{}_\sigma,
\end{equation*}
and $\eps^{\mu\nu\rho\sigma}$ is the 4 dimensional Levi-Civita pseudo-tensor. Define the dual tensor ${}^*F{}^{\mu\nu}:= \frac{1}{2}\eps{}^{\mu\nu\rho\sigma}F{}_{\rho\sigma}$. Projecting as $b{}^\mu=u{}_\nu {}^*F{}^{\mu\nu}$ leads to\footnote{Assuming ideal MHD, in the comoving fluid-frame, the electric field vanishes: $u{}^\mu F{}_{\mu\nu}=0$.}\footnote{A factor of $\sqrt{4\pi}$ has been absorbed into the magnetic field definition throughout.}:
\begin{align*}
  b{}^0 &= \frac{W \mc{B}{}^i v{}_i}{\alpha}, &
  b{}^i &=  \frac{\mc{B}{}^i + \alpha b{}^0 W \tilde v{}^i}{W}, &
  b{}^2 &= g{}_{\mu\nu} b{}^\mu b{}^\nu = \frac{(\alpha b{}^0)^2 + \mc{B}{}^i \mc{B}{}^j \gamma{}_{ij}}{W^2}.
\end{align*}

We form evolution equations for MHD as a balance (conservation) law as in~\cite{Banyuls:1997zz}:
\be
\label{eq:claw}
\pd_t \mathbf{q} + \pd_i \mathbf{F}^i = \mathbf{s},
\ee
where the conservative variables $\mathbf{q}$ are associated to the Eulerian observer. In our convention $\mathbf{q} := (D, S_j, \tau, B^k)$. Note that we take as primitive variables $\mathbf{w} = (\rho, \tilde u^i, p, B^i)$ and may map to $\mathbf{q}$ through:
\begin{equation*}
  \mathbf{q} = \sqrt{\gamma}
  \left(
    \rho W,
    (\rho h + b^2) W \tilde u{}_j - \alpha b{}^0 b{}_j,
    (\rho h + b^2) W^2 - \rho W - \left(p + \frac{b^2}{2}\right)
    - (\alpha b{}^0)^2, \mc{B}{}^k
  \right).
\end{equation*}
The inversion of this map, where required, is evaluated numerically (\S\ref{ssec:constoprim}).

In order to determine the flux vector $\mathbf{F}{}^i$, and source $\mathbf{s}$ of Eq.\eqref{eq:claw} one considers the conservation of rest mass density $\nabla{}_\mu[\rho u{}^\mu]=0$, the Bianchi identity $\nabla{}_\mu[T{}^{\mu\nu}]=0$, and the Maxwell relation $\nabla{}_\mu[{}^*F{}^{\mu\nu}]=0$. The former two equations are projected onto and normal to $\Sigma_t$ whereas the latter is projected onto $\Sigma_t$. This results in:
\begin{equation}
  \label{eq:fluxB_def}
  \mathbf{F}^i = 
  \begin{bmatrix}
  &D \alpha \tilde v^i  \\
  & S_j \alpha \tilde v^i + \delta^i_j \left(p + \frac{b^2}{2}\right) \alpha \sqrt{\gamma} - \frac{\alpha \sqrt{\gamma} b_j \mc{B}^i}{W}\\
  & \tau \alpha \tilde v^i + \alpha \sqrt{\gamma} \left(\left(p + \frac{b^2}{2}\right) v^i  - \frac{\alpha b^0 \mc{B}^i}{W}\right) \\
  & \alpha (B^k \tilde v^i - B^i \tilde v^k)
  \end{bmatrix}.
\end{equation}
and:
\begin{equation*}
  \label{eq:hydsrc_def}
  \mathbf{s} = \alpha \sqrt{\gamma}
  \begin{bmatrix}
  & 0   \\
  & T^{00}\left(\half \beta^i\beta^k \pd_j \gamma_{ik} - \alpha \pd_j \alpha  \right) + T^{0i} \beta^k \pd_j \gamma_{ik} + T^0_i \pd_j \beta^i + \half T^{ik}\pd_j \gamma_{ik}\\
  & T^{00}\left(\beta^i\beta^j K_{ij} - \beta^i \pd_i \alpha  \right) + T^{0i} \left(2 \beta^j K_{ij} - \pd_i \alpha\right) +  T^{ij}K_{ij} \\
  & 0^k   
  \end{bmatrix}.
\end{equation*}
One further projection can be formed. The Maxwell relation when projected in the normal direction to $\Sigma_t$ gives rise to the divergence-free constraint on $B{}^i$, that is $\partial{}_i [B{}^i] = 0$. This will be automatically enforced by the discretisation of the equations (\S\ref{ssec:constrained_transport}). 
Finally, the MHD equations are coupled to the \z4c{} system (\ref{sssec:z4c_system}) through the the ADM sources (here prefixed with superscript $A$):
\begin{equation}
\begin{gathered}
\begin{aligned}
  {}^A\rho =&  (\rho h + b^2) W^2 - \left(p + \frac{b^2}{2}\right) - (\alpha b{}^0)^2,\\
  {}^A S{}_{i} =&  (\rho h + b^2) W^2 v{}_i - b{}^0 b{}_i \alpha,\\
  {}^A S{}_{ij} =&  (\rho h + b^2) W^2 v{}_i v{}_j + \left(p+ \frac{b^2}{2}\right) \gamma{}_{ij} - b{}_i b{}_j.
\end{aligned}
\end{gathered}
\label{eq:admmatcoupling}
\end{equation}

\subsection{Equation of state}
The GRMHD system of conservation laws together with the magnetic field divergence-free condition restricts $8$ degrees of freedom for the $9$ primitive variables $(\rho,p,\epsilon,\tilde u^i, B^i)$. The role of the equation of state (EOS) is to reduce this underdeterminedness through imposition of a thermodynamical relationship.

Here we highlight two classes of EOS used in \GRAthena{}:
\begin{enumerate}
  \item[i)]{Ideal gas with Gamma-law:
    A relation between $(\rho, p, \epsilon)$ is formed with $p=\rho \epsilon(\Gamma-1)$ imposed where $\Gamma=1+1/n$ and $n$ is the adiabatic index of the fluid\footnote{Henceforth we restrict to the $\Gamma=2$ case.}. During initialisation this is supplemented by the barotropic relation $p=K\rho^\Gamma$ where $K$ serves as a problem-dependent, fluid mass-scale.
  }
  \item[ii)]{Tabulated, 3-dimensional finite-temperature:
    First, species fractions $Y_i$ that track fluid composition are introduced (CMA scheme \cite{Plewa:1998nma}). For simplicity, consider only the electron fraction $Y_\mathrm{e}$. This extra variable is incorporated into the GRMHD system as an additional evolved variable $D Y_e$ with zero source. Overall this allows us to write the relation $p = p\left( \rho, T, Y_\mathrm{e} \right)$. Quantities $\rho$ and $T$ are tabulated in logarithmic space, and, together with $Y_\mathrm{e}$ are linearly interpolated so as to provide $\log p$ when required.

    Our implementation supports tables in the form used by the CompOSE database \cite{Typel:2013rza} processed     to also include the sound speed by the PyCompOSE tool \footnote{https://bitbucket.org/dradice/pycompose/}.
  }
\end{enumerate}

\subsection{Numerical technique}\label{ssec:num_tech}
In treatment of the GRMHD system in \GRAthena{} the method of lines approach is taken. In this overview attention is restricted to time-evolution based on the third order, three stage, low-storage SSPRK$(3,3)$ method of \cite{gottlieb2009highorderstrong}.

Field components involved in the \z4c{} sub-system (\S\ref{sssec:z4c_system}) are sampled on vertex-centers (VC). Spatial derivatives on the interior of the computational domain $\mathrm{int}(\Omega)$ (i.e.~in the bulk away from $\partial\Omega$) are computed with high-order, centered, finite difference (FD) stencils with one-point lop-siding for advective terms (see e.g.\cite{Brugmann:2008zz}). The ghost-layer extent (choice of $\Ng$) fixes the overall, formal FD approximation order as $2(\Ng-1)$ throughout $\mathrm{int}(\Omega)$. The interpolation order during calculation of $\mathcal{P}$ under mesh refinement is consistently selected to match that of the FD; whereas $\mathcal{R}$ is exact for VC.

To fix physical boundary conditions, within every time-integrator sub-step, an initial Lagrange extrapolation (involving $\Ng+1$ points) is performed so as to populate a ghost-layer at $\partial\Omega$. The dynamical \z4c{} equations then furnish data for $\{\chi,\,\tilde{\gamma}{}_{ij},\,\alpha,\,\beta{}^i\}$ at $\partial\Omega$  whereas values there for $\{\hat{K},\,\tilde{\Gamma},\,\Theta,\,\tilde{A}{}_{ij}\}$ are based on the Sommerfeld prescription of \cite{Hilditch:2012fp}.

During evolution the Courant-Friedrich-Lewy (CFL) condition must be satisfied. In the context of \GRoAthena{} grid-spacing on the most refined level of a refinement hierarchy determines the global time-step that is applied on each \MeshBlock{}. Finally for the evolved \z4c{} state-vector, high-order Kreiss-Oliger (KO) dissipation \cite{Kreiss:1973}, of degree $2\Ng{}$, with uniform factor $\KO{}$ is applied.

In the case of the MHD sub-system (\S\ref{sssec:mhd_system}) a standard conservative, Godunov-style finite volume method is employed. Thus, practically, when the conserved variables $\mathbf{q}$ are to be evolved, we require fluxes (Eq.\eqref{eq:fluxB_def}) evaluated at cell interfaces. Interface data is prepared based on reconstruction of known cell-averaged quantities to the left and right of an interface. Once prepared we can utilize e.g.~Local-Lax-Friedrichs (LLF):
\begin{equation*}
  \hat{\mathbf{F}}_{i+1/2}=\frac{1}{2}\left(
    \mathbf{F}[\mathbf{q}^-_{i+1/2}] +
    \mathbf{F}[\mathbf{q}^+_{i-1/2}] -
    \alpha (\mathbf{q}^+_{i-1/2} - \mathbf{q}^-_{i+1/2})
  \right),
\end{equation*}
where $\alpha$ is taken as the maximum over characteristic speeds, which, are computed from locally reconstructed variables, while we note the availability of 
more sophisticated Riemann solvers for magnetic fields such as those presented in \cite{Kiuchi:2022ubj}. In the absence of magnetic fields exact characteristic speeds are provided in \cite{Banyuls:1997zz} whereas for MHD we follow the approximate prescription of \cite{Gammie:2003rj}. In the MHD context preservation of the magnetic field divergence-free condition requires some care. This is achieved through a constrained transport scheme which we return to in \S\ref{ssec:constrained_transport}.

\subsubsection{Intergrid transfer, and reconstruction}\label{ssec:intergrid}

In \GRoAthena{}, data transfer between differing discretizations for the evolved field components of GR and MHD is required at various points of the time-evolution algorithm. Indeed CC grids form the primary representation for the evolved, conserved quantities $\mathbf{q}$, furthermore the FC and EC grids are utilized for constrained transport. The (smooth) geometric data is evolved on VC. For simplicity, this latter, is transferred to the other grids, as required, at linear order.

In order to evaluate the numerical fluxes at the cell interfaces, we reconstruct the primitive variables $\mathbf{w}$ from CC to FC. Reconstruction avoids the increase of total variation, providing a limited interpolation strategy appropriate for functions of locally reduced smoothness.
To this end the Piecewise Linear Method (PLM) \cite{vanLeer:1974a}, and Piecewise Parabolic Method (PPM) \cite{Colella:2011a,McCorquodale:2015a}, as implemented in \Athena{}, are further supplemented in \GRAthena{} by WENO5 \cite{Jiang:1996a} and WENOZ \cite{Borges:2008a}.

\subsubsection{Constrained transport}\label{ssec:constrained_transport}

The ideal MHD approximation reduces the Maxwell equations to a hyperbolic equation for the magnetic field components,
the induction equation, and an elliptic equation, Gauss's Law. The solution of this elliptic equation at each time-step
in a numerical simulation through e.g. a relaxation method would be incredibly costly and so various schemes have been developed
to avoid this requirement, as discussed above.  
\Athena employs an implementation of the constrained transport approach to ensure divergence free evolution of the magnetic field, 
following the original proposal of \cite{Evans:1988a}, with the development of the specific implementation in 
\Athena following \cite{Gardiner:2005,Gardiner:2007nc,Stone:2008mh,Beckwith:2011iy,White:2015omx}.

Here the magnetic field is discretised, not as a cell average as for the other hydrodynamical variables,
or as in other implementations such as the Flux-CT implementation of \cite{Toth:2000}, but as face averages:
\begin{eqnarray}
	B^x_{i-1/2,j,k} = \frac{1}{\Delta A}\int_{A_{i-1/2,j,k}} B^x \mathrm{d}y \mathrm{d}z,
\end{eqnarray}
with the other components $B^y,B^z$ discretised on, respectively, the $y$ and $z$ faces of the computational cell
$A_{i,j-1/2,k}$, $A_{i,j,k-1/2}$, with $\Delta A$ the area of the face being integrated over.

Noting that the electric field is given by the cross product of the magnetic field with the fluid velocity $\alpha \tilde v$,
integrating the induction equation over a face of the computational cell gives an evolution equation 
for the face averaged magnetic field in terms of the integrated curl of the electric field over the face.
By applying Stokes theorem, the face centred magnetic field is updated by evaluation of the integral of the 
edge centred, time averaged, electric field around the face in question:
\begin{eqnarray}
	\int^{t_{n+1}}_{t_n} \partial_t B^{x}_{i-1/2,j,k} \mathrm{d}t &=& \frac{1}{\Delta A}\int^{t^{n+1}}_{t^n} \mathrm{d}t
  \int_{A_{i-1/2,j,k}} \Big\{
  - \partial_y (\alpha(B^x  \tilde v^y - B^y  \tilde v^x)) \nonumber\\
  &&
  - \partial_z (\alpha(B^x \tilde v^z - B^z \tilde v^x))
  \mathrm{d}y \mathrm{d}z\Big\}, \nonumber\\
	&=& \frac{1}{\Delta A}\int^{t^{n+1}}_{t^n} \mathrm{d}t \int_{A_{i-1/2,j,k}} -\partial_y (E^z) + \partial_z (E^y) \mathrm{d}y \mathrm{d}z, \nonumber\\
	B^{x,n+1}_{i-1/2,j,k}	&=& B^{x,n}_{i-1/2,j,k} -\frac{\Delta t}{\Delta A} \int_{\partial A_{i-1/2,j,k}} \mathbf{E}^{n+1/2} \cdot \mathbf{\mathrm{d}l},
\end{eqnarray}
where $\mathbf{\mathrm{d}l}$ is the line element around the edge of face $A_{i-1/2,j,k}$, and $\mathbf{E}^{n+1/2}$ the time averaged electric field between $t_n$ and $t_{n+1}$.

It is clear that this discretisation of the field automatically preserves the divergence of $B$ between time-steps;
when evaluating the divergence of $B^{n+1}$ using the above expression, contributions from the electric field will cancel
due to shared edges between faces having opposite signs due to the orientations of the integrals around the face edges.
The divergence of $B^{n+1}$ must then equal the divergence of $B^n$. 

The final requirement for evaluating the magnetic field is to calculate the time averaged, edge centred, electric fields.

Following the standard Godunov approach, the time averaged face centred electric field can be calculated as the solution to the Riemann 
problem at the cell interface, which is calculated along with the hydrodynamical fluxes. Then the cell centred electric field 
can be explicitly constructed as the cross product of the magnetic field and fluid 3-velocity. The derivative of the electric field from the cell centre to face 
centre can be calculated, and then upwinded to the cell face in the upwind direction as determined by the sign of the mass flux. This gradient is then used to integrate
the time averaged face centred electric field to the cell edge.

A detailed description of this procedure may be found in \cite{White:2015omx}, though we note that the implementation for dynamical
space-times requires the appropriate densitisation of variables by the time dependent metric determinant, which can no longer be
factored out of time derivatives.

For simulations with AMR new \MeshBlock{}s will be created dynamically, on which data will be populated through interpolation from the 
coarser (finer) \MeshBlock{}(s) that they replace. Naively, this interpolation will not respect the divergence free constraint on $B$,
and so \Athena implements the restriction and prolongation operators of \cite{Toth:2002a} designed to preserve the curl and divergence 
of interpolated vector fields.

At the interface between refinement levels, faces and edges of \MeshBlock{}s of different levels overlap. To ensure a conservative evolution
the face centred hydrodynamical fluxes are kept consistent, by replacing the face centred flux on the coarser block with the area weighted 
sum of the fluxes on the finer blocks. Similarly for the magnetic field evolution, the edge centred electric field on the coarser block
is replaced with the length weighted sum of the edge centred fields on the finer block. To ameliorate other effects such as 
the non-deterministic order of {\tt MPI} processes, shared edge centred electric fields are replaced by their averages also in the case of neighboring 
\MeshBlock{}s on the same refinement level.

\subsection{Conservative-to-primitive variable inversion}\label{ssec:constoprim}
In order to evaluate the fluxes in Eq.~\eqref{eq:claw} we convert the conservative variables, $\mathbf{q}$ to their primitive representation $\mathbf{w}$ by inverting the definitions of $\mathbf{q}$ in Eq.~\eqref{eq:fluxB_def}. This involves solution of a non-linear system for the primitive variables\footnote{A variety of strategies are summarised and compared in \cite{Siegel:2017sav}.}. To this end \GRAthena{} features a custom, error-policy based implementation \PrimitiveSolver{}. We have also coupled the external library \reprimand{}\footnote{We find robust performance when employing a tolerance of $10^{-10}$ in the
bracketing algorithm and a ceiling of $20$ on the velocity variable $\tilde
u$.} of \cite{Kastaun:2020uxr}. Here we note some additional considerations related to the latter method. Generically the conservative-to-primitive inversion is evaluated pointwise, where required, over a {\tt MeshBlock}, though is not subject to the underlying particulars in choice of discretization.

Physically we expect a neutron-star (NS) to have a well defined, sharp surface. Consider a radial slice of the fluid density from the NS center: a non-negative profile that drops to zero is expected. Resolving such sharp features can lead to numerical issues during the variable conversion process. In order to address this we follow the standard technique of introducing a tenuous atmosphere region where primitive variables are set according to:
\begin{align}
  \rho_{\mathrm{atm}} &= f_{\mathrm{atm}} \rho_{\mathrm{max}}, &
  p_{\mathrm{atm}} &= p(\rho_{\mathrm{atm}}), &
  \tilde u^i_{\mathrm{atm}} &= 0;
\end{align}
with $\rho_{\mathrm{max}}$ the maximum value of the fluid density in
the initial data, and $f_{\mathrm{atm}}$ a problem-specific parameter. The magnetic field is left unaltered in the atmosphere. During the course of the evolution, physical processes of the star, or numerical dissipation, may drive the fluid into the atmosphere region. On account of this we impose an additional criterion where if $\rho<\rho_{\mathrm{thr}} = f_{\mathrm{thr}} \rho_{\mathrm{atm}}$ the cell is reset to atmosphere\footnote{Typically we set $f_{\mathrm{thr}} = 1$ with $f_{\mathrm{atm}}\in[10^{-13},\,10^{-18}]$.}. Similarly cells are reset to atmosphere if conversion from conservative to primitive variables fails to converge. When atmosphere values are set, we recalculate
the conservative variables from the atmosphere primitives and metric variables.

\subsection{Diagnostic quantities}
For later convenience common scalar diagnostic quantities computed during simulations are collected here.

\subsubsection{Energetics}\label{sssec:diag_energetics}
The baryon mass is a conserved quantity defined as:
\be\label{eq:Mb}
M_b = \int_{\Sigma_t} D \d^3 x;
\ee
where the integral is evaluated over the entire computational domain. Various energy contributions to the GRMHD system are similarly evaluated:
\begin{align}
  E_{\rm Kin} &= \frac{1}{2} \int_{\Sigma_t} \frac{S_iS^i}{D} \d^3 x, &
  E_B &= \frac{1}{2}\int_{\Sigma_t} \sqrt{\gamma}W b^2\d^3 x, &
  E_{\rm int} &= \int_{\Sigma_t} D\epsilon \d^3 x.
\end{align}

\subsubsection{Gravitational wave extraction}\label{sssec:wave_extraction}
To obtain the gravitational wave (GW) content of the space-time, we calculate the Weyl scalar $\Psi_4$,
the projection of the Weyl tensor onto an appropriately chosen null tetrad $k,l,m, \bar{m}$ with the conventions of \cite{Brugmann:2008zz}. Multipolar decomposition onto spin-weighted spherical harmonics\footnote{Convention here is that of \cite{Goldberg:1966uu} up to a Condon-Shortley phase factor of $(-1)^m$. } ${}_s Y_{lm}$ of spin-weight $s=-2$ is evaluated through:
\begin{eqnarray}\label{eq:ipswsh}
  \psi_{\ell m} &=& \int^{2\pi}_0\int^\pi_0 \Psi_4\,\bar{{}_{-2}Y_{\ell m}}(\vartheta,\,\varphi)\sin \vartheta \mathrm{d}\vartheta \mathrm{d} \varphi.
\end{eqnarray}
Thus $\Psi_4$ is first calculated at all grid points throughout the \Mesh{} whereupon data is interpolated onto a set of geodesic spheres $Q_R$ (see \S\ref{ssec:geodesic_spheres}) at given extraction radii $R_Q$. Subsequently the numerical quadrature of Eq.\eqref{eq:ipswsh} may be evaluated. In order to select the grid level parameter $n_Q$ that controls the number of samples on $Q_R$ we may match the area element to that of any intersected \MeshBlock{} objects through $n_Q = \left \lceil \sqrt{\frac{\pi R_Q^2}{\sp^2} - 2} \right \rceil$.
The modes of the GW strain $h$ may be computed from the projected Weyl scalar by integrating twice in time $\psi_{\ell m}=\ddot{h}_{\ell m}$.
The strain is then given by the mode-sum:
\begin{equation}
  h_+ - i h_\times = \frac{G}{D_L} \sum_{\ell=2}^{\infty}\sum_{m=-\l}^\l h_{\ell m}(t)\; {}_{-2}Y_{\ell m}(\vartheta,\varphi)\,,
\end{equation}
where $D_L=(1+z)R$ is the luminosity distance of a source located at distance $R$ to the observer and at redshift $z$. Phase and frequency conventions follow the LIGO algorithms library \cite{lalsuite}.

\subsection{Task-list execution model}\label{ssec:task_list}

\begin{figure}[ht!]
	\centering
  \includegraphics[width=0.85\columnwidth]{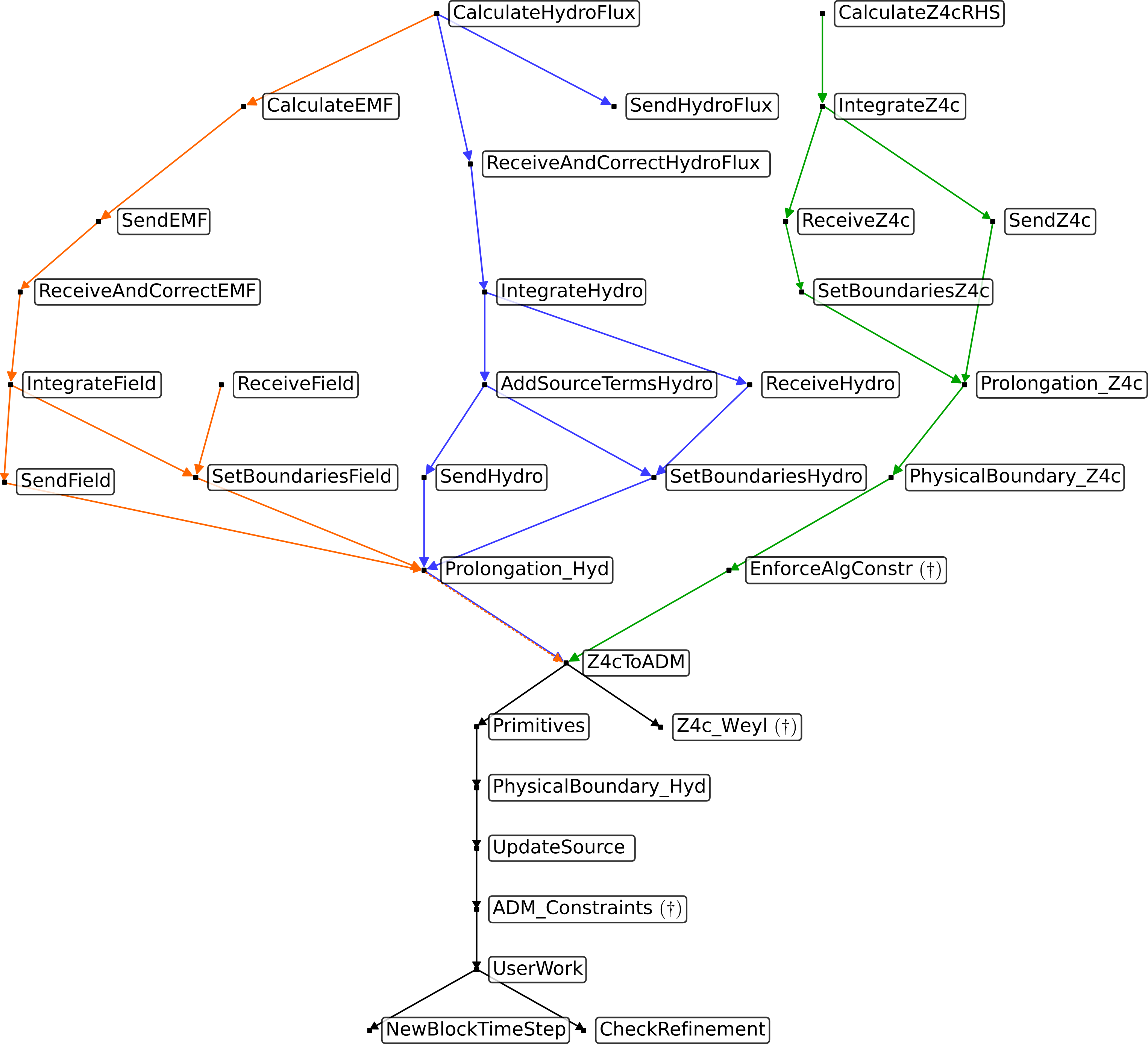}
  \caption{%
    Simplified DAG of GRMHD task-list in \GRAthena{} executed on each {\tt MeshBlock} during a time-integrator sub-step. Tasks are boxed and arrows between nodes indicate direction of dependency. In the absence of a dependency, remaining tasks may be executed in parallel. Labels indicate operation being performed. Arrows are color-coded according to type of field data manipulated where: (orange) corresponds to magnetic field; (blue) fluid; (green) \z4c{}; (black) general operations. Tasks labeled with a $(\dagger)$ are only executed on the final integrator sub-step.    }
\label{fig:tasklist_hydro}
\end{figure}

The individual components of a calculation during a time-evolution sub-step that must be executed must satisfy a particular order of execution. The task-based execution model employed in \GRoAthena{} may be viewed as assembling this information in a directed, acyclic graph (DAG). A task specifies a given operation that must be performed and may feature dependency on the prior completion of other tasks. These may be assembled so as to represent a more complicated calculation into a so-called task-list. This allows to neatly encapsulate the calculation steps involved in GRMHD in Fig.\ref{fig:tasklist_hydro}.
One advantage of this form of calculation assembly is that wait time during buffer-communication can be masked as e.g.~system calculations proceed in parallel, this feature improves efficiency. Thus, for example, as shown in Fig.\ref{fig:tasklist_hydro} once the {\tt CalculateZ4cRHS} and {\tt IntegrateZ4c} tasks are completed on the current {\MeshBlock} a suitable subset of this data may be sent via communication buffers ({\tt SendZ4c} task) to a nearest-neighbor {\tt MeshBlock} so as to populate its ghost-layer (and vice-versa through the {\tt RecvZ4c} task). Once these tasks complete, boundary data on the current {\tt MeshBlock} may be set as represented by {\tt SetBoundariesZ4c} and prolongated where required. As a final note, operations involving global data reduction (e.g.~numerical quadrature over geodesic sphere in gravitational wave calculation \S\ref{ssec:geodesic_spheres}) are performed external to the task-list.

\section{Applications}
We now demonstrate the performance and accuracy of \GRAthena{} through solution of problems in dynamical space-times with GRMHD. A subset of tests based on our work \cite{Cook:2023bag} is summarised for NS space-times, followed by a set of first applications, in the long term study of magnetic field development in an isolated NS, and in using AMR to efficiently study the Kelvin-Helmholtz instability during a BNS merger.

Throughout this discussion $\Ng{}=4$ is selected which sets $6^\mathrm{th}$ order accurate finite differencing in treatment of \z4c{}. Damping parameters are taken as $\kappa_1=0.02$, and $\kappa_2=0$. Kreiss-Oliger dissipation is fixed at $\KO{}=0.5$.

\subsection{Single Star Spacetimes} \label{sec:ss}
\subsubsection{GRHD evolution of Static Neutron Star }\label{sec:ss:sns_grhd}

Our first test is of the evolution of a static NS model known as A0 in the convention of \cite{Dimmelmeier:2005zk},
commonly used as a code benchmark, \eg~\cite{Font:2001ew,Thierfelder:2011yi}.
The star has baryon mass $M_b=1.506\Mo$ and gravitational mass $M=1.4\Mo$, with initial data calculated by solving the Tolman-Oppenheimer-Volkoff
(TOV) equations using a central rest-mass density of 
$\rho_c=7.905\times10^{14}\gccm$, with an ideal gas EOS with $\Gamma=2$, and 
initial pressure set using the polytropic EOS with $K=100$.

Our grid configuration has outer boundaries at $[\pm 378.0,\pm 378.0,\pm 378.0]$~km, with 6 levels of static mesh refinement fully covering the star, with the innermost level covering the region $[\pm 14.8, \pm 14.8, \pm 14.8]$~km.
The grid resolution is set through the {\tt Mesh} parameter $N_M=32,48,64,128$, which gives a grid spacing of $[369,246,185,92.3]$~m on the finest level.
\begin{figure}[h]
  \centering 
    \includegraphics[width=0.6\textwidth]{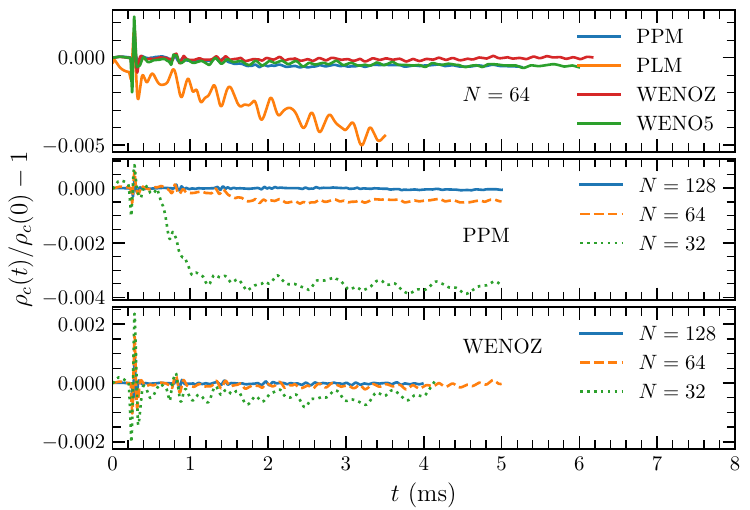}
  \caption{GRHD Evolution of central rest-mass density of the static
      star A0 model. Top: varying reconstruction at constant resolution
      $N_M=64$. Central, Bottom: varying resolution for PPM and WENOZ reconstruction respectively.
      Note: Some data has been truncated to $5$~ms for ease of visualisation.
    (Figure adapted from \cite{Cook:2023bag}).
      }
    \label{fig:A0_rhoc}
\end{figure}
\begin{figure}[h]
  \centering 
    \includegraphics[width=0.6\textwidth]{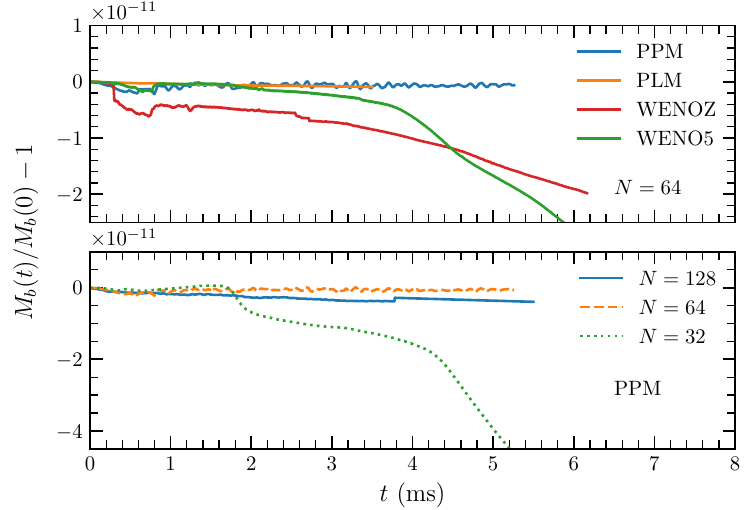}
  \caption{Baryon mass conservation during evolution of the static star A0 model. The mass conservation can be violated due to the
      artificial low-density atmosphere as the fluid expands beyond
      the computational domain.
    (Figure adapted from \cite{Cook:2023bag}).
      }
    \label{fig:A0_masscons}
\end{figure}
We first demonstrate the evolution of the central rest mass density of the star $\rho_c(t)$ as a function of resolution and for a variety of reconstruction schemes in Fig.~\ref{fig:A0_rhoc}.  The star is perturbed by the presence of truncation errors, leading to the excitation of characteristic oscillations in the central density, the frequencies of which match those predicted by perturbation theory for the first three radial modes, at frequencies of $\nu_F=1462,3938$ and $5928$~Hz \cite{Baiotti:2008nf}, and the relative amplitudes of which remain below $10^{-2}$, and converge away as resolution increases.
We find that, as expected, PLM is the least accurate scheme, with PPM performing poorly at low resolutions, but performing consistently with WENO schemes at higher resolutions.

In Fig.~\ref{fig:A0_masscons} we demonstrate a relative conservation of baryon mass at the level of $10^{-11}$ consistently across schemes and resolutions, with violations occuring due the setting of atmosphere levels and matter leaving the computational domain. The best performance is found for PPM, which preserves the star surface most accurately.

\begin{figure}[h]
  \centering 
    \includegraphics[width=0.6\textwidth]{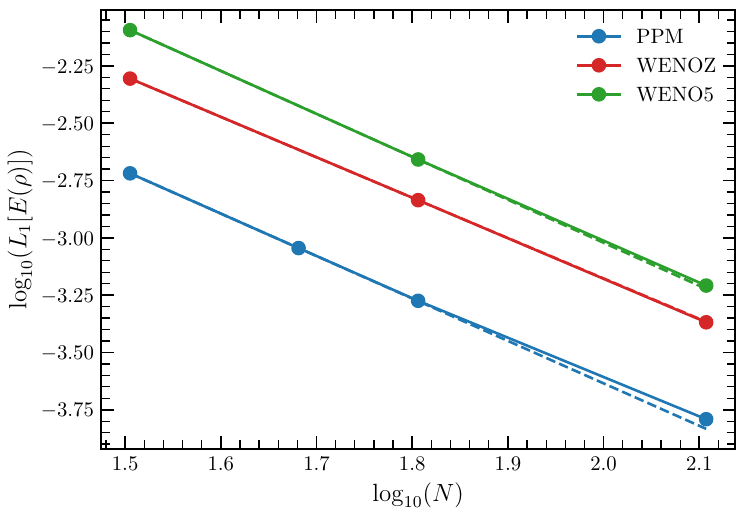}
  \caption{Convergence of $L_1[E(\rho)]:=|\rho(\mathbf{x},t) - \rho(\mathbf{x},0)|$ for static star model A0 
  at $t=2.96$ ms.
      Order of convergence is $1.85,1.76,1.87$ for PPM, WENOZ, WENO5 reconstruction schemes respectively.
  (Figure adapted from \cite{Cook:2023bag}).
    }
 \label{fig:A0_L1rho}
\end{figure}
\begin{figure}[h]
  \centering 
    \includegraphics[width=0.6\textwidth]{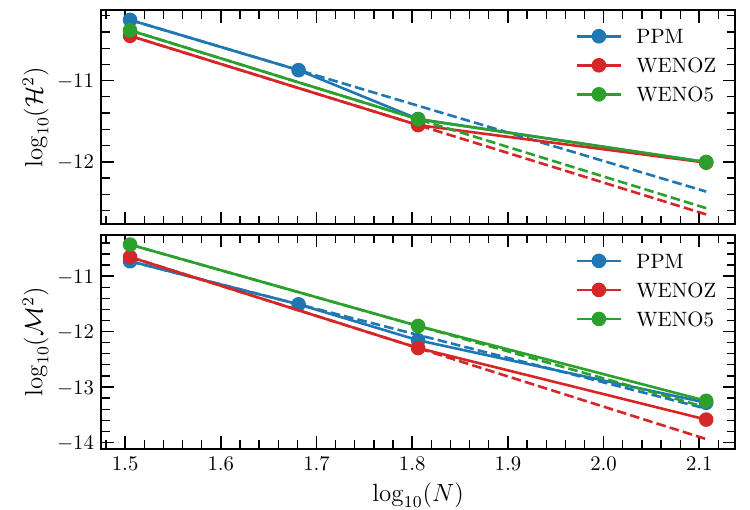}
  \caption{Hamiltonian (top) and momentum (bottom) constraints for static star model A0 
  at $t=2.96$ ms.
      Order of convergence in the Hamiltonian (momentum) constraint is 
      $3.51 (4.41), 3.65 (5.46), 3.64 (4.88)$ for PPM, WENOZ, WENO5 respectively.
      (Figure adapted from \cite{Cook:2023bag}).
    }
 \label{fig:A0_ham}
\end{figure}
The convergence of three, spatially integrated, quantities is demonstrated for this model: the absolute difference between the density profile at time $t$ and the initial density profile; and the Hamiltonian and Momentum constraints (\S\ref{sssec:z4c_system}). We show the former in Fig.~\ref{fig:A0_L1rho}, with dashed lines showing linear extrapolations from the first two data points, the slopes of which indicate the order of convergence. Approximately $\2nd{}$ convergence is found, most clearly seen in the WENO(5,Z) schemes, while at the resolutions shown the PPM scheme has the lowest overall error. In Fig.~\ref{fig:A0_ham}, the constraints are shown to converge at an order between 3.51 and 5.46 depending on the reconstruction scheme used.

 \subsubsection{Long-term Magnetic field dynamics in Static Neutron Star}\label{sec:ss:sns_grhmd}

We now present an application of \GRAthena{} to the long term evolution of magnetic fields within
NSs. The existence of a magnetic field configuration that is stable over long timescales within
a NS is an open problem, with analytic and numerical studies showing the instability of purely poloidal 
fields on Alfven timescales, and suggesting the need for the development of a toroidal component 
to stabilise the field \cite{Tayler:1957a,Tayler:1973a,Wright:1973a,Markey:1973a,Markey:1974a,Flowers:1977a,Braithwaite:2005md},
with long term numerical relativity evolutions necessary to understand the rearrangement of the field profile
and energy distribution \cite{Kiuchi:2008ss,Lasky:2011un,Ciolfi:2011xa,Pili:2017yxd,Sur:2021awe}.

We evolve an isolated star given by model A0, using WENOZ reconstruction, the grid set up of \S\ref{sec:ss:sns_grhd} and resolutions $N_M=32,64,128$, with a purely poloidal field given by ~\cite{Liu:2008xy}:
\begin{subequations}
\begin{align}
        &(A_x,A_y,A_z) = (-yA_\varphi, x A_\varphi, 0), \\
        &A_\varphi := A_b \max(p-0.04\, p_{\rm max},0). \label{eq:Apotential}
\end{align}
\end{subequations}
where $p_{\mathrm{max}}$ is the maximum value of the pressure within
the star. 

The magnetic field is given by $ B = \nabla \times A$, with $A_b$ a free parameter scaling 
the overall strength of the field. Below we set this so that the maximum value of the magnetic 
field inside the star is $1.84\times10^{16}$~G, approximately the magnetic field strength expected 
in magnetars.

We compare our results with our earlier simulations of a similar system, performed in the Cowling approximation in \cite{Sur:2021awe}.

\begin{figure}[h]
  \centering 
    \includegraphics[width=0.6\textwidth]{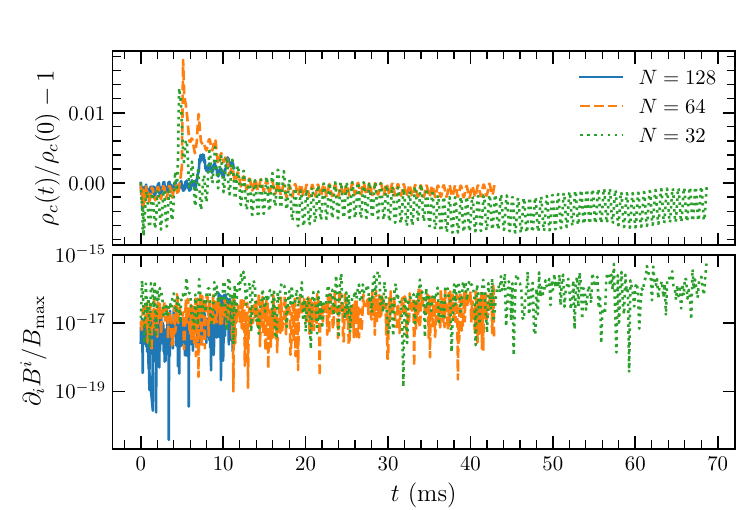} 
  \caption{Top: Evolution of central rest-mass density of the static
      star A0 model with superposed poloidal magnetic field, and WENOZ reconstruction.
      Bottom: $\partial_i B^i$ integrated over full computational domain $\Omega$ and normalised by $B_{\mathrm{max}}:=\max_{\Omega} B$.
    (Figure adapted from \cite{Cook:2023bag}).
     }
 \label{fig:A0_mhd_rhoc_divb}
\end{figure}
\begin{figure}[h]
  \centering 
    \includegraphics[width=0.6\textwidth]{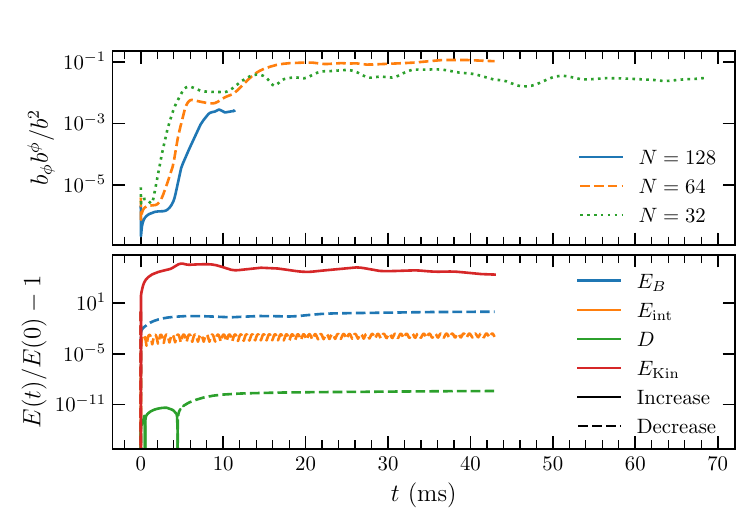} 
  \caption{Top: Growth of the toroidal component of magnetic field energy 
     as a fraction of the total for model shown in Fig.\ref{fig:A0_mhd_rhoc_divb}. Bottom: relative change in 
     magnetic energy, internal energy, mass and kinetic energy for $N_M=64$. Solid lines denote 
     a positive change in energy, dashed lines denote a negative change.
    (Figure adapted from \cite{Cook:2023bag}).
     }
 \label{fig:A0_mhd_b2_E}
\end{figure}
We first demonstrate the oscillations of the central density of the star in the upper panel of Fig.~\ref{fig:A0_mhd_rhoc_divb}. In comparison to the case in the absence of magnetic fields (Fig.~\ref{fig:A0_rhoc}) we see larger oscillations, driven by the superposed $B$-field. We see the peak in oscillations corresponds to the saturation time of the toroidal field compoenent growth.
Throughout the evolution the constrained transport algorithm maintains the divergence free condition on the magnetic field with a relative violation
of ${{\sim}10^{-16}}$ shown in the lower panel of Fig.~\ref{fig:A0_mhd_rhoc_divb}.

As expected we see a growth of a toroidal field component to stabilise the poloidal component, shown in the upper panel of Fig.~\ref{fig:A0_mhd_b2_E}. After $\sim{}16$ ms, the ratio of toroidal 
to total magnetic field energy $b^\phi b_\phi / b^\mu b_\mu$ grows to $\sim{}10 \%$, with the saturation of this growth delayed as resolution increases. This result is in line with the Cowling simulations demonstrated in Fig.~3 of \cite{Sur:2021awe}.

We present the energy budget of the space-time in the lower panel of Fig.~\ref{fig:A0_mhd_b2_E} with energies defined in \S\ref{sssec:diag_energetics}. As the magnetic field profile rearranges it drives fluid motion leading to a growth in kinetic energy, with local maxima in the kinetic energy profile tracking the maxima within the toroidal field strength. In agreement with Fig.~5 of \cite{Sur:2021awe} we see a loss in overall magnetic field energy, though we see a much improved retention of internal energy, which we attribute to an improved atmosphere treatment, and the increased grid size in comparison to the previous simulations.

\begin{figure*}[t]
   \centering 
     \includegraphics[width=0.32\textwidth]{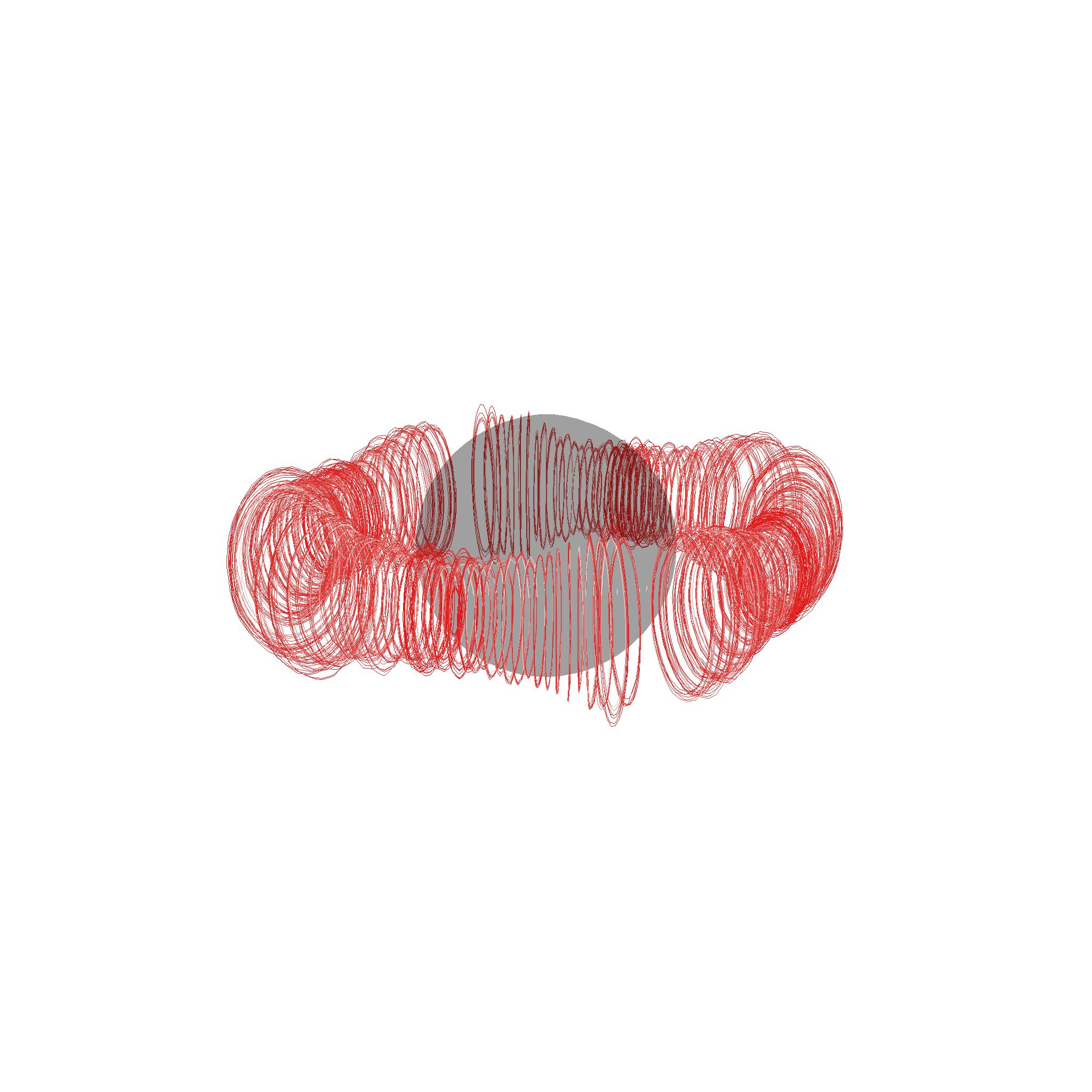}
     \includegraphics[width=0.32\textwidth]{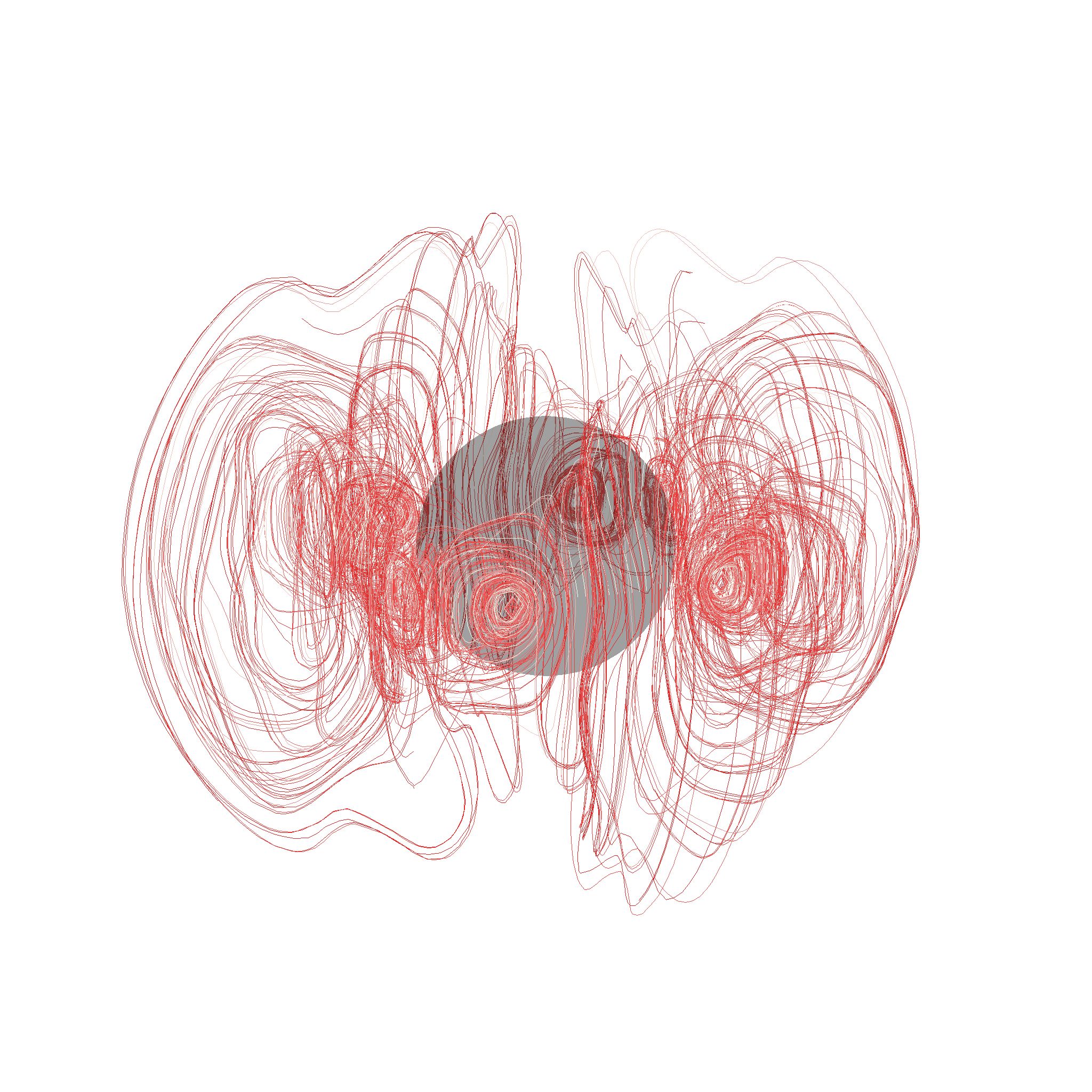}
     \includegraphics[width=0.32\textwidth]{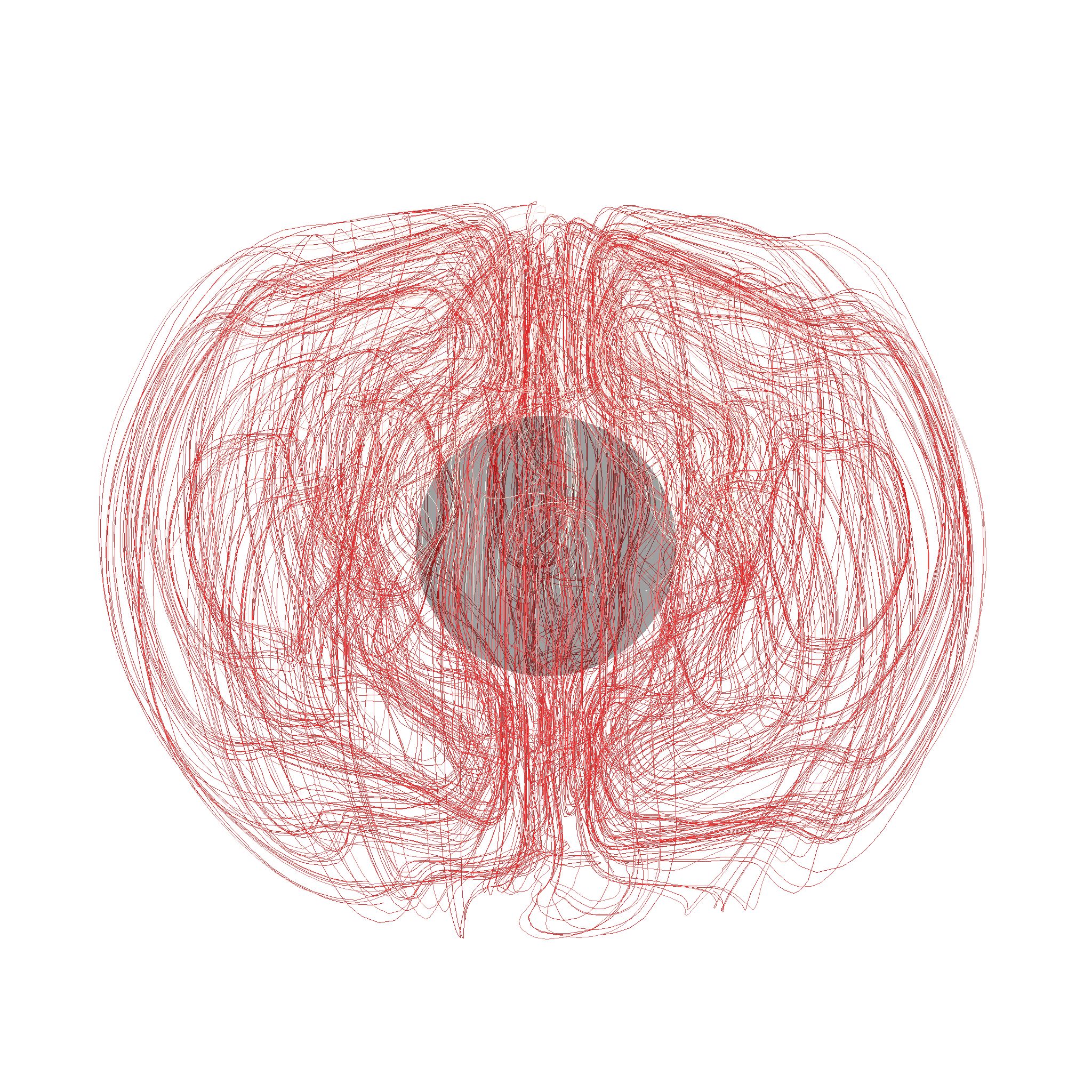}
     \caption{Evolution of magnetic streamlines seeded in equatorial plane at
       radius $5.91$ km. Gray isocontour shows surface of $\rho = 6.18\times 10^{14} \gccm$.
       Left: Varicose instability at $t=4.93$~ms.
       Middle: Kink instability at $t=12.8$~ms.
       Right: Late time non-linear field arrangement $t=44.3$~ms.
      (Figure adapted from \cite{Cook:2023bag}).
       }
  \label{fig:A0_mhd_streamline}
 \end{figure*}

Through $3$D visualisations (Fig.~\ref{fig:A0_mhd_streamline}) of the magnetic field lines seeded on a circle in the equatorial plane of radius $5.91$ km, we observe the onset of the ``varicose'' and ``kink'' instabilities of \cite{Tayler:1957a}. In a cylindrical system these correspond to, respectively, an $m=0$ and $m=1$ perturbation. The first of these breaks the rotational invariance of the fieldlines, with the initially constant cross-sectional area now varying as a function of the azimuthal angle around the star, with the latter displacing the fieldlines in the direction perpendicular to the gravitational field. 

After $4.93$ ms (left panel) the cross sectional area of the streamlines is no longer rotationally invariant, a sign of the onset of the varicose instability, with the kink instability disrupting streamlines orthogonal to the equatorial plane by $12.8$ ms (middle panel), coinciding with the saturation of the growth of the toroidal field component. After $44.3$ ms (right panel) we see an overall poloidal structure with clear toroidal contributions after the non linear dynamical evolution of the star.

\subsection{Gravitational collapse of Rotating Neutron Star}\label{sec:collapse}

The ability of \GRAthena{} to handle gravitational collapse is confirmed, by simulating the unstable, uniformly rotating, equilibrium model D4 of \cite{Baiotti:2004wn,Reisswig:2012nc,Dietrich:2014wja}. This model is triggered to collapse by the introduction of a perturbation to the pressure field of 0.5$\%$, following \cite{Dietrich:2014wja}. The initial data is calculated with the \texttt{RNS} code of \cite{Stergioulas:1994ea} with a central density of $\rho_c = 1.924388\times10^{15}\gccm$ and a polar-to-equatorial coordinate axis ratio of $r_p/r_e = 0.65$. This leads to a star with rotational frequency of $1276$~Hz, baryon mass of $M_b=2.0443\Mo$ and gravitational mass $M=1.8604\Mo$. 

Our grid consists of 8 refinement levels, 3 of which lie within the star radius, with the outer boundaries the same as for \S\ref{sec:ss:sns_grhd}, without imposed grid symmetries. We perform simulations with a base level resolution $N_M=32, 64, 128$ corresponding to  finest grid spacings of $92.2,46.1,23.1$~m respectively, using PPM reconstruction. The evolution successfully passes through gravitational collapse and horizon formation without the need for excision.
\begin{figure}[ht]
  \centering 
    \includegraphics[width=0.6\textwidth]{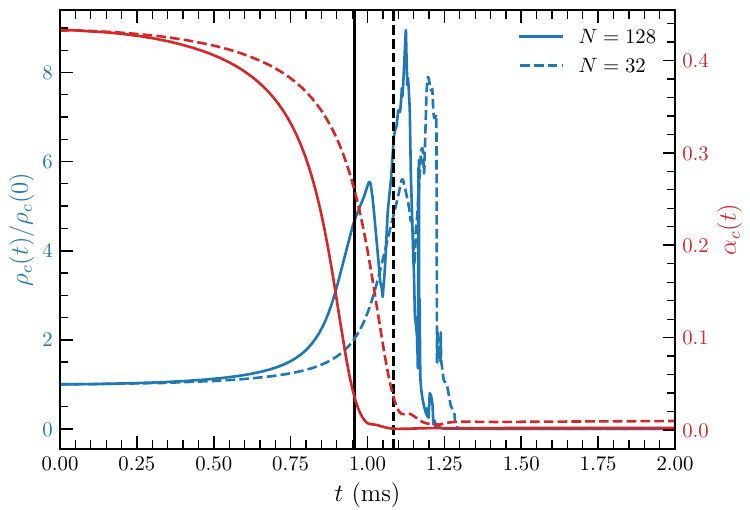}
  \caption{Collapsing D4 model for simulations wth PPM reconstruction. Evolution of central rest-mass density and lapse of at
      different resolutions. Vertical black lines denote collapse
      time for each simulation. Oscillatory behaviour occurs 
  within the horizon, after collapse.
    (Figure adapted from \cite{Cook:2023bag}).
    }
    \label{fig:D4_rhoc_alpc}
\end{figure}
\begin{figure}[ht]
  \centering 
  \includegraphics[width=0.6\textwidth]{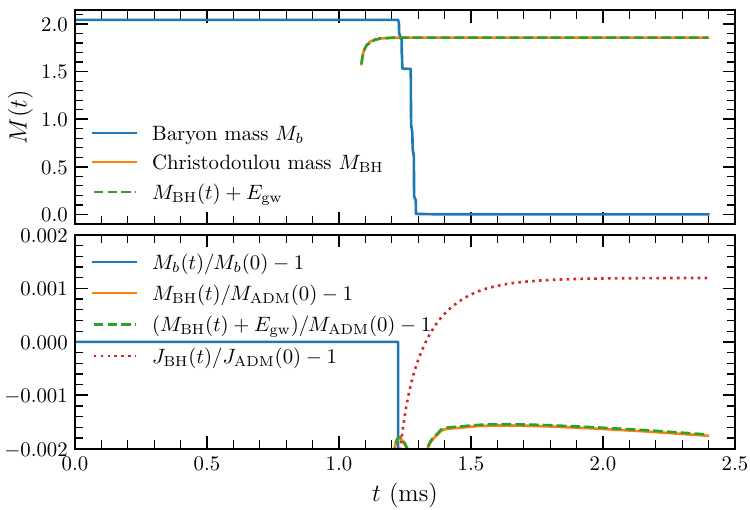}
  \caption{Collapsing D4 model for simulations wth PPM reconstruction. Top: Evolution of baryon mass and black hole (Christodoulou) mass for $N_M=32$. The dashed line shows the
    sum of the black hole mass and the gravitational-wave energy in
    the $(2,0)$ mode extracted at the coordinate sphere.
    Bottom: Baryon mass and angular momentum conservation and
    relative differences of Christodoulou mass with and without gravitational-wave energy added with respect to
    the initial ADM mass.
    (Figure adapted from \cite{Cook:2023bag}).
    }
    \label{fig:D4_masses}
\end{figure}

We first monitor the central density and lapse as indicators of the star collapse (Fig.~\ref{fig:D4_rhoc_alpc}). As the density and hence curvature increase at the centre of the star, the moving puncture gauge conditions used here \cite{Thierfelder:2010dv}, cause the lapse to collapse at the origin, and handle black hole formation.

Using an apparent horizon finder, employing the spectral fast-flow algorithm of \cite{Gundlach:1997us,Alcubierre:1998rq} we locate an apparent horizon (AH) after one ms, at which point we see losses of baryon mass from the grid  (Fig.~\ref{fig:D4_masses}) as the determinant of the spatial metric becomes very large at the centre of collapse. Prior to this we see mass  conservation with relative error $10^{-13}$ despite the presence of refinement boundaries within the star, providing a robust test of the flux correction of \Athena{}. From the AH shape we calculate the Christodoulou mass \cite{Christodoulou:1970wf} of $M_{\rm BH}=1.857\Mo$ and angular momentum $J_{\rm BH}=1.884\Mo^2$G/c of the horizon. These values differ from the expected ADM mass and angular momentum of the space-time by a relative error of order $10^{-3}$ for the lowest resolution simulation $N_M=32$, when the energy loss through GWs is included, which improves as a function of resolution. These results favourably compare to previous tests for this model in \cite{Reisswig:2012nc,Dietrich:2014wja}.

\begin{figure}[t]
  \sidecaption
  \centering
  \includegraphics[width=0.60\textwidth]{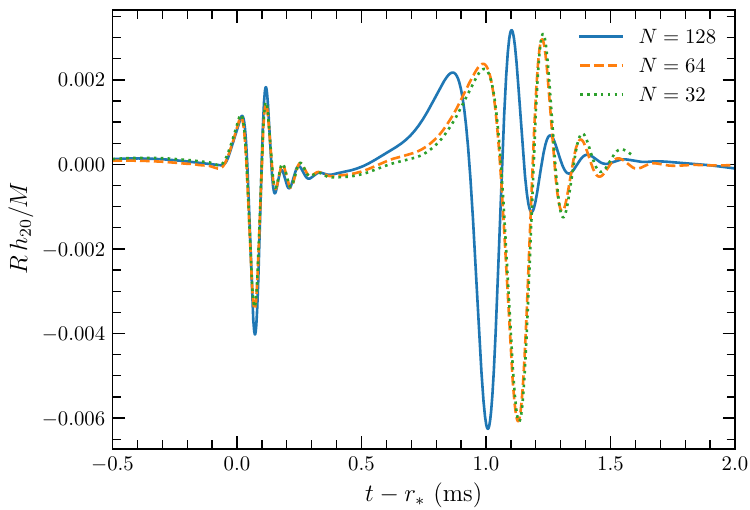}
  \caption{Dominant $(2,0)$ multipole of the gravitational strain for the collapsing D4 model for simulations wth PPM reconstruction.
                At $N_M=32$ the resolution is too low to observe convergent behaviour, though at higher resolutions we do see convergence in the waveforms at approximate first order.
Note usage of geometric units on the vertical axis.
    (Figure adapted from \cite{Cook:2023bag}).
  }
  \label{fig:D4_waves}
\end{figure}

Gravitational waveforms (\S\ref{sssec:wave_extraction}) are extracted using geodesic spheres at radius $R=221$~km, and the dominant $(\ell,m)=(2,0)$ of the strain is computed integrating in time and adjusting for drift \cite{Baiotti:2008nf}.
In Fig~\ref{fig:D4_waves} we demonstrate these as a function of the retarded time to the extraction spheres, given by $t-r_*$ with  $r_* = r + 2 M \log(r/2M- 1)$, and $r(R)$ the areal radius of the spheres of coordinate radius $R$ (the isotropic Schwarzschild radius).

The waveform can be characterised by an initial burst of unphysical radiation after which we see the ``precursor-burst-ringdown'' behaviour expected for gravitational collapse. At $t-r_*\simeq 1$~ms for run $N_M=128$ we see a local maximum in GW amplitude, corresponding to the formation of the AH, followed shortly after by the global maximum amplitude after collapse,  with the quasi-normal mode oscillations of a perturbed Kerr BH forming the final part of the wave signal.

\subsection{Binary Neutron Star space-times}\label{sec:bns}

The evolution of BNS space-times, and the extraction of highly accurate gravitational waveforms is a key target for \GRAthena{}. We demonstrate this capability through direct, cross-code comparison of BNS evolution, without magnetic fields (GRHD), against the \BAM{} code \cite{Bruegmann:1996kz,Brugmann:2008zz,Thierfelder:2011yi}.
Self consistency of \GRAthena{} is established through self-convergence of gravitational waveforms, together with good mass conservation for both GRHD and GRMHD simulations.

The initial data evolved is irrotational and constraint satisfying, as generated by the \texttt{Lorene} library \cite{Gourgoulhon:2000nn}, describing a quasi-circular equal-mass merger with baryon mass $M_b=3.2500\Mo$ and gravitational mass $M=3.0297\Mo$ at an initial separation of $45$~km. The ADM mass of the binary is $M_{\rm ADM}=2.9984\Mo$, the angular momentum $J_{\rm ADM}=8.83542\Mo^2$G/c, and the initial orbital frequency is $f_0\simeq294$~Hz.

An ideal gas EOS is used with $K=124$ and, in the case in which magnetic fields are present they are initialised as in the single star case with the vector potential in Eq.~\eqref{eq:Apotential}, with $A_b$ chosen to give a maximum value of $1.77\times 10^{15} G$.

For runs with magnetic fields the grid outer boundaries are set at $[\pm 2268$, $\pm 2268$, $\pm2268]$~km with no symmetry imposed. A refined static {\tt Mesh} is defined with 7 levels of refinement, with the innermost grid covering both stars, from $[\pm 37, \pm 37, \pm37]$~km. For runs without magnetic fields the grid is identical
except for the imposition of bitant symmetry across the plane $z=0$.
The {\tt Mesh} parameters are $N_M=64,96,128$, corresponding to a grid spacing on the finest refinement level of $(554,369,277)$~m respectively. Simulations are performed with WENOZ reconstruction, and hydrodynamical variables are excised within the apparent horizon for runs with GRMHD.

\subsubsection{Benchmark against \BAM} \label{sec:bns:vsBAM}

\begin{figure*}[ht]
  \centering
  \includegraphics[width=0.56\textwidth]{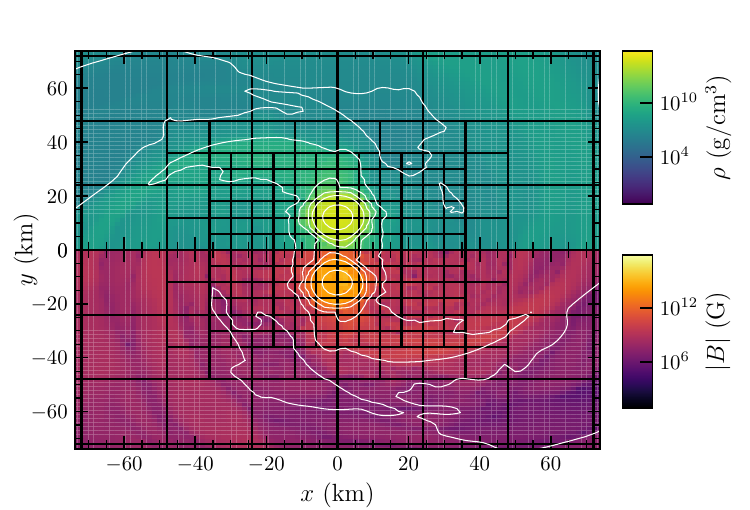}
  \includegraphics[width=0.56\textwidth]{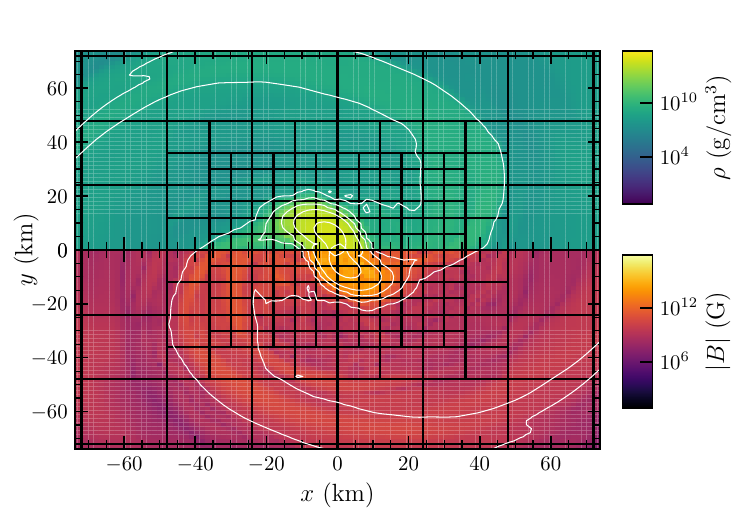}
  \includegraphics[width=0.56\textwidth]{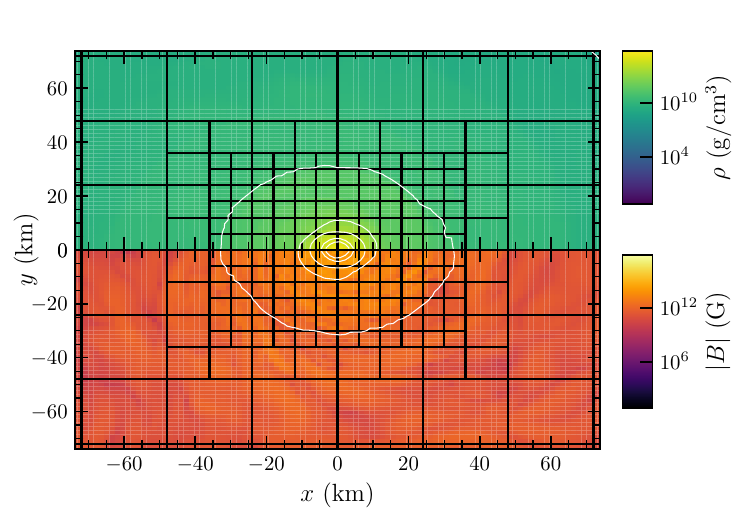}
    \caption{Snapshots of rest-mass density (upper half) and magnetic field strength
      (lower half)
      in the orbital plane. (Top, middle, bottom) correspond to respectively the final orbit ($t=5.96$ms), the moment of merger ($t=7.83$ms),
      and a late stage shortly before gravitational collapse $t=16$ms. White contours are lines of constant density at values $\rho =(6.18 \times 10^{14}, 3.09\times 10^{14}, 6.18 \times 10^{13}, 6.18 \times 10^{12},6.18 \times 10^{10}, 6.18 \times 10^{8}, 6.18 \times 10^{6}, 6.18 \times 10^{4}) \gccm$.
    (Figure adapted from \cite{Cook:2023bag}).
      }
    \label{fig:bns_snap2d}
\end{figure*}

From the beginning of the simulation with GRHD, the binary inspiral lasts for $\sim3$ orbits before merging to form a massive remnant star that undergoes gravitational collapse at ${\sim}19$~ms for the lowest resolution simulation. In the case of GRMHD  we see similar behaviour in the inspiral phase and visualise the the pre-merger, merger and pre-collapse phases in Fig.~\ref{fig:bns_snap2d}.

We now directly compare to evolutions in \BAM{}.
\GRAthena{} and \BAM{} employ slightly different hydrodynamical evolution schemes. For instance, the latter reconstructs the primitives using the fluid internal energy, rather than pressure as in \GRAthena{}. \BAM{} also lacks the intergrid operators of \GRAthena{} (\S\ref{ssec:intergrid}), and uses a box-in-box style refinement strategy with Berger-Oliger time sub-cycling. For the sake of comparison, we perform two \BAM{} simulations with differing hydrodynamical schemes, one using the LLF flux and WENOZ reconstruction similar to \GRAthena{}, and one using the entropy flux-limited (EFL) scheme of \cite{Doulis:2022vkx}, which is a fifth-order accurate scheme.
The \BAM{} grid uses six refinement levels, two of which move and contain the stars. The finest refinement level entirely covers the star at a resolution of ${\sim} 461$~m, that is comparable to the \GRAthena{} grid configuration with $N_M=64$.

\begin{figure}[h]
  \centering
    \includegraphics[width=0.6\textwidth]{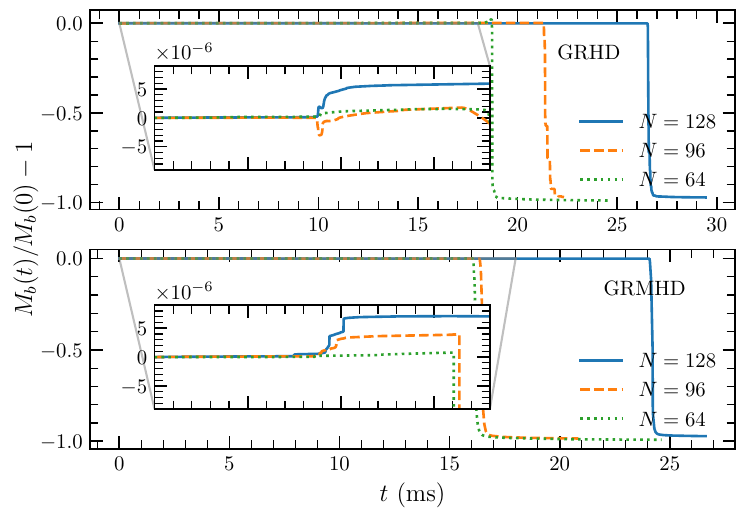}
  \caption{%
    Baryon mass conservation for GRHD and GRMHD evolutions of the BNS. To the moment of merger, $t\simeq7.5$~ms, the maximum relative violation is of the order $10^{-7}$. After merger, the violation increases to the $10^{-6}$  level. Note the mass typically increases due to atmosphere accretion. The abrupt drop of the mass at late times is due to black hole formation. %
        (Figure adapted from \cite{Cook:2023bag}).
}
\label{fig:BNS_mass}
\end{figure}

The conservation of baryon mass is depicted in Fig.~\ref{fig:BNS_mass} for the GRHD and GRMHD evolutions at different resolutions. The maximum violation of the relative conservation ${\sim}10^{-6}$ occurs at the lowest resolution. This error, while larger than for the single star, is sufficiently low to accurately study ejecta properties, and is consistent with other state-of-the-art Eulerian codes at the considered resolutions,
e.g.~\cite{Radice:2018pdn}. It was found that through varying atmosphere parameter settings those leading to a mass increase provide the best mass conservation.
\begin{figure}[h]
  \centering
  \includegraphics[width=0.6\textwidth]{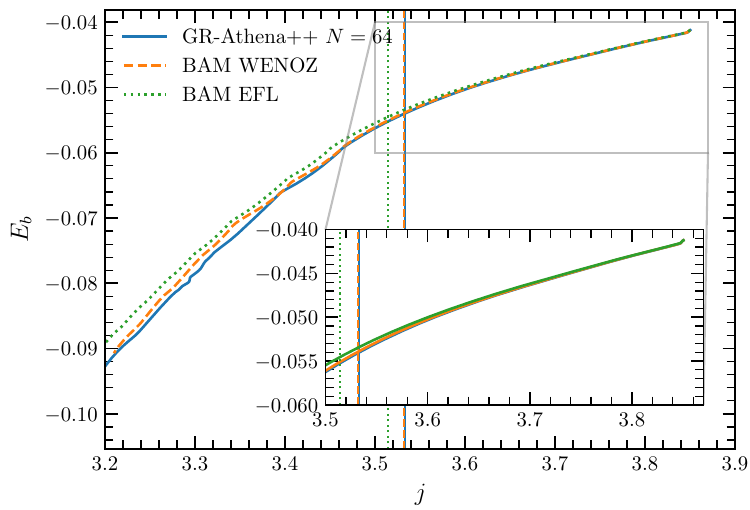}
  \caption{%
    Binding energy \textit{vs} angular momentum curves $E_b(j)$ for \GRAthena{} and \BAM{} simulations. These quantities are computed following \cite{Damour:2011fu,Bernuzzi:2012ci}; $E_b := (M_{\mathrm{ADM}} - E_{\mathrm{GW}})\nu/(M-1) $, $j := (J_{\mathrm{ADM}} - J_{\mathrm{GW}})/(M^2\nu)$, with $\nu$ the symmetric mass ratio, $\nu := q/(1+q)^2$, $q>1$ the mass ratio and quantities suffixed with $\mathrm{GW}$ the contributions radiated away in gravitational waves.
    The moment of merger of each dataset is shown as vertical line. The inset captures energetics during the orbital phase, prior to the moment of merger.
        (Figure adapted from \cite{Cook:2023bag}).
}
\label{fig:BNS_Ej}
\end{figure}

To compare evolutions in a gauge invariant manner, curves of binding energy against angular momentum \cite{Damour:2011fu,Bernuzzi:2012ci}, are considered in Fig.~\ref{fig:BNS_Ej}. Close agreement up to merger at relatively low resolutions is found, with differences more pronounced post-merger. We note that the differences between \GRAthena{} and \BAM{} are consistent with the internal differences between evolution schemes in \BAM{}.
\begin{figure*}[t]
  \centering
  \includegraphics[width=0.99\textwidth]{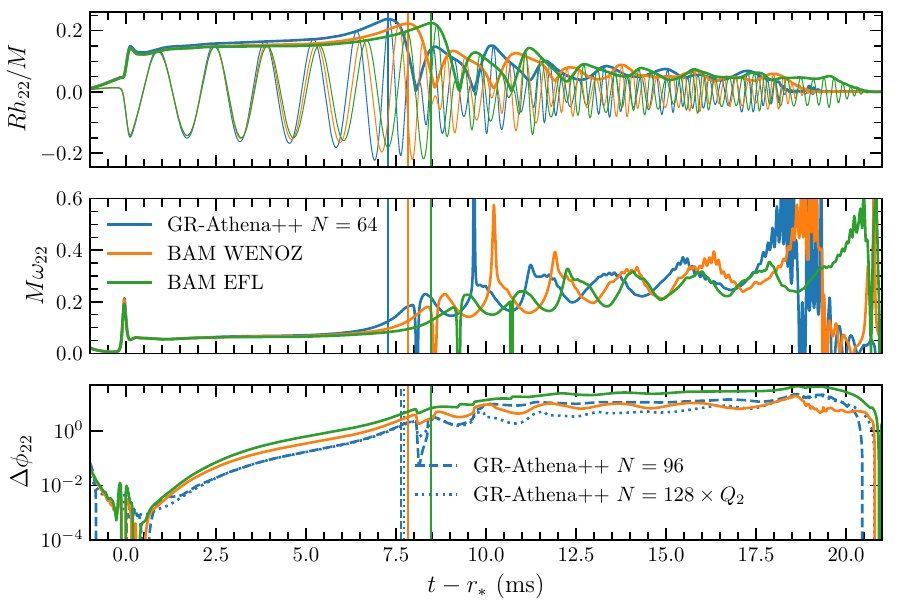}
  \caption{Waveform analysis and comparison with \BAM{}.
    (Top): Real part and amplitude of strain mode $h{}_{22}$. We show: \GRAthena{} data with WENOZ reconstruction and $N_M=64$; \BAM{} data with WENOZ reconstruction, and the EFL scheme at a similar resolution.
    (Middle): Instantaneous frequency of the waves for the three datasets.
    Note that in the top and middle panels the vertical axis is in geometric units.
  (Bottom): Phase differences between the \GRAthena{} data at $N_M=64$ and: $N_M=96$ (dashed); $N_M=128$ (dotted lines); and \BAM{} data. \GRAthena{} phase difference data at $N_M=128$ is rescaled assuming $\2nd{}$ convergence, matching with the $N_M=96$ data up to merger. The moment of merger of each dataset is indicated with a vertical line.
    (Figure adapted from \cite{Cook:2023bag}).
  }
\label{fig:BNS_wvf}
\end{figure*}

In Fig.~\ref{fig:BNS_wvf}, direct comparison of $h{}_{22}$ between the three runs is made. The top panel compares the real part of the gravitational wave strain, where we see amplitudes closely matching, and the main difference coming in the phasing.
The earlier merger time\footnote{Merger time is signalled by the amplitude peak of $h{}_{22}$. } of \GRAthena{} suggests that this run is slightly more dissipative than its \BAM{} counterpart. Collapse time of runs is consistent to within $2.5$ ms.

The middle panel demonstrates the instantaneous gravitational wave frequency, which closely match at the moment of merger, $\omega_{22}\simeq 1340$~Hz for \GRAthena{} and \BAM{} WENOZ and $\omega_{22}\simeq 1360$~Hz for \BAM{} EFL. Similar consistency between runs is found in the frequency evolution of the post-merger remnant up to collapse, after which the quasi-normal modes of the remnant BH are insufficiently resolved, but are compatible with the fundamental mode frequency $\nu_{\rm QNM}\sim6.5$~kHz ($M\omega_{22}\simeq0.57$ in geometric units).

In the bottom panel of Fig.~\ref{fig:BNS_wvf}, phase differences of the \GRAthena{} $N_M=64$ run with respect to the \BAM{} runs and the \GRAthena{} runs at higher resolutions are quantified. The difference between the $N_M=64$ and $N_M=128$ runs is rescaled by $Q_2$ to compare to the difference between the $N_M=64$ and $N_M=96$ run assuming $\2nd{}$ order convergence\footnote{Here $Q_n:=(\delta x_c^n - \delta x_m^n)/(\delta x_c^n - \delta x_f^n)$ where $\sp_c$, $\sp_m$, and $\sp_f$ are coarse, medium, and fine grid-spacings.}. These curves closely match up to merger suggesting $\2nd{}$ order convergence during the inspiral phase. Further, the difference to the \BAM{} runs significantly reduces at higher resolutions.

\subsubsection{Kelvin-Helmholtz instability}\label{sec:bns:KH}

The Kelvin-Helmholtz instability (KHI) arises over a shearing interface in a fluid with a discontinuous velocity profile. This interface is unstable to perturbation, breaking down and forming small scale vortex like structures. It is suggested \cite{Price:2006fi} that this should be expected during the
merger of a BNS, as the two stars first make contact they shear past each other, creating an unstable interface. In the presence of magnetic fields numerical studies \cite{Kiuchi:2015sga,Kiuchi:2017zzg,Palenzuela:2021gdo,Aguilera-Miret:2023qih} support that KHI leads to magnetic field amplification.

As small scale vortices must be resolved, efficiency considerations motivate the use of targeted AMR. Here, two AMR criteria are demonstrated to capture $B$-amplification. We perform two SMR based simulations denoted SMR64 and SMR128, with the same grid configuration as those in \S\ref{sec:bns} utilising {\tt Mesh} sampling $N_M=64$ and $N_M=128$.
The AMRB and AMR$\sigma$ runs are initialised from SMR64 after $t=4.93$ms, allowing an extra refinement level to be generated according to the appropriate criterion, bringing the maximum refinement up to that of run SMR128. Grid resolution is bounded below by that of SMR64.

In the case of AMRB, a \MeshBlock{} (MB) is refined if its maximum value of $|B|$ exceeds $8.35 \times 10^{13}$G and derefined if lower than this value. For AMR$\sigma$, a MB is refined if % its maximum value of
$\mathrm{max}_{\Omega_{i\in Z}}\sigma > 4.50 \times 10 ^9 \mathrm{cm}/\mathrm{s}$, and derefined if $\mathrm{max}_{\Omega_{i\in Z}}\sigma < 2.25 \times 10 ^9 \mathrm{cm}/\mathrm{s}$,
where $\sigma := \sqrt{((\Delta_x \tilde u^y)^2 + (\Delta_y \tilde u^x)^2)}$
represents the shear of the fluid velocity
and $\Delta_i$ is the undivided difference operator in the $i$th direction.

\begin{figure*}[ht]
   \centering
   \includegraphics[width=0.9\textwidth]{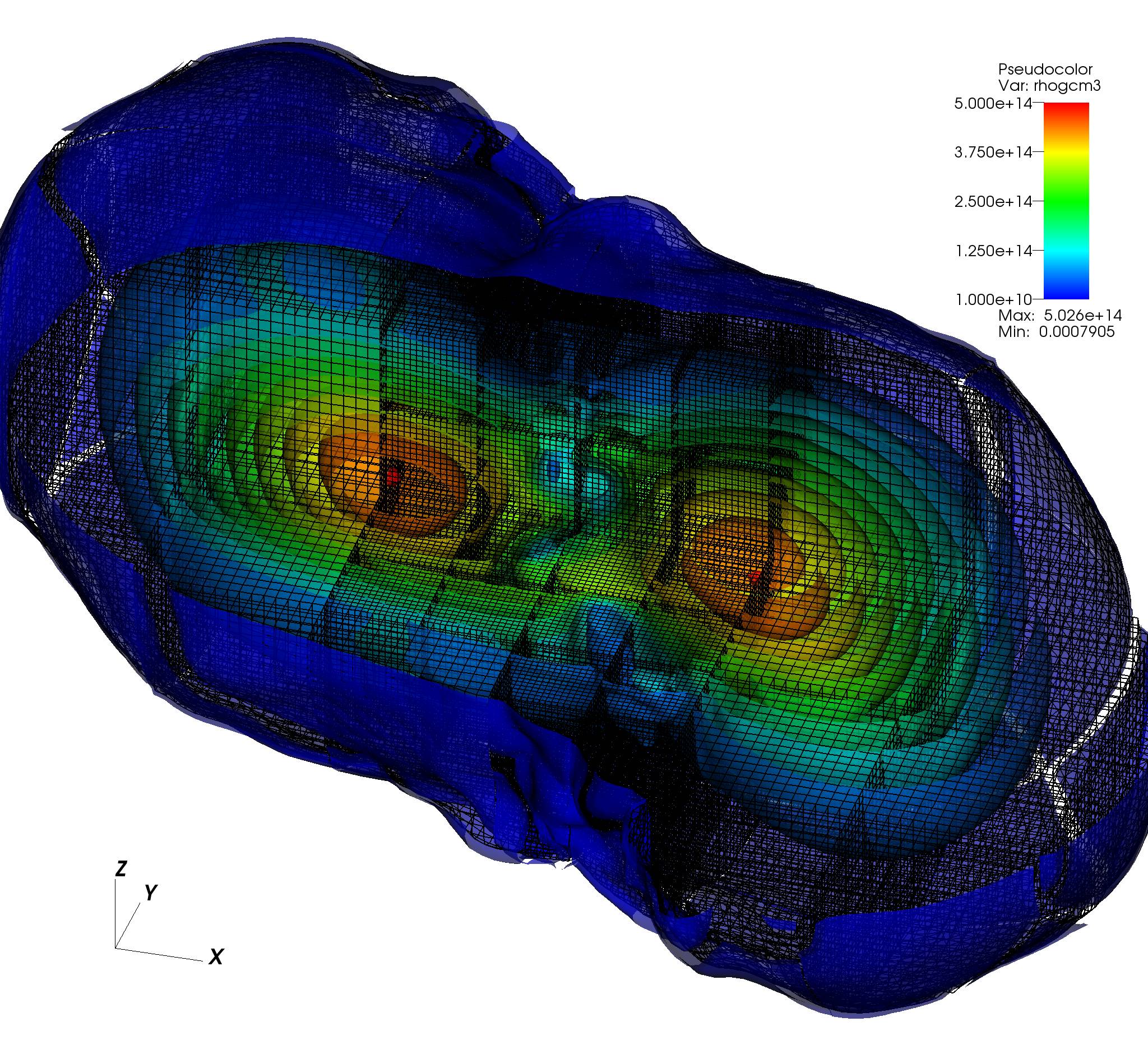}
   \caption{\Mesh{} structure at $t=7.88$ms shortly before merger in the AMR$\sigma$ run. Isocontours of density are shown for $\rho > 10^{11} \gccm$. To reveal internal structure the $z>0,~y<0$ quadrant is suppressed.
      (Figure adapted from \cite{Cook:2023bag}).
    }
  \label{fig:BNS_KHI_AMRgrid}
 \end{figure*}

The grid structure of run AMR$\sigma$ at $t=7.88\,\mathrm{ms}$, shortly before merger, is shown in Fig.~\ref{fig:BNS_KHI_AMRgrid}, where small MBs generated at the star interface can be seen.

\begin{figure}[h]
  \sidecaption
	\centering
  \includegraphics[width=7.2cm]{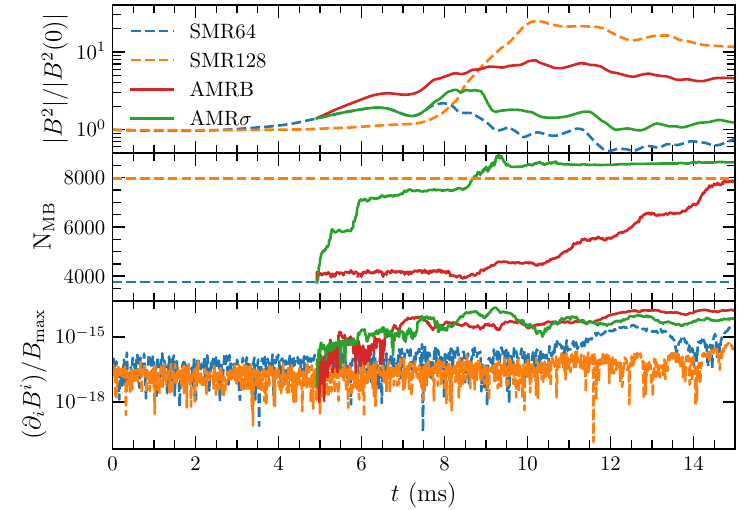}
  \caption{%
  Top: Amplification of total magnetic field strength $|B^2|$ in BNS merger for SMR, AMRB, and AMR$\sigma$ refinement.
    Middle: Total $\#\mathrm{MB}$ present in each run. %
  Bottom: $\partial_i B^i$ integrated over full computational domain $\Omega$ and normalised by $B_{\mathrm{max}}:=\max_{\Omega} B$.
    (Figure adapted from \cite{Cook:2023bag}).
  }
  \label{fig:BNS_KHI}
\end{figure}

The performance of an AMR criterion is judged on its ability to capture $B$-amplification, the $\#\mathrm{MB}$ generated (as a measure of efficiency), and the violation of the divergence free constraint of the magnetic field introduced through MB creation and destruction operations, demonstrated in the upper, middle and lower panels of Fig.~\ref{fig:BNS_KHI} respectively.
In the upper panel we see that in SMR128 the energy is amplified by a factor of over 25 after the moment of merger, at $t\sim10.2~$ms. The AMR simulations capture less of this amplification, with an AMRB amplification factor of 7.85, while AMR$\sigma$ only captures a factor of 3.26. In \cite{Kiuchi:2015sga}, at considerably higher resolutions of $17.5$m a much larger amplification, of 6 orders of magnitude, from a much weaker initial magnetic field profile, of initial strength order $10^{13}$G is observed. Our expectation is that through addition of further levels of mesh refinement we should be able to capture such larger amplifications.

In the middle panel, we see that up to the moment of the peak magnetic field energy the run AMRB generates less than $10\%$ more MBs than SMR64, a factor of 1.79 fewer than SMR128 and a factor 1.91 fewer at the moment of merger. This supports the idea that targeted use of AMR can resolve magnetic field amplification. AMR$\sigma$ clearly however generates more MB than SMR128. This is due to MBs being generated at the star surface and in the region of unphysical ejecta from the merger. This criterion performs worse at capturing $B$-amplification, but shows improved mass conservation, which may have applications in ejecta tracking for simulations featuring physical ejecta.

Finally in the lower panel we see that the relative violation of the divergence-free constraint, as integrated over the domain, remains below $10^{-14}$ for the AMR runs.
Note that, while the overall $\#\mathrm{MB}$ in the middle panel appears approximately constant, oscillations indicate continual destruction and creation of \MeshBlock{} objects. Crucially, divergence preserving interpolation \cite{Toth:2002a}, ensures an absence of any secular trends in $\partial_i B^i$ post AMR activation.

\subsection{Scaling tests}\label{sec:scaling}
The (binary, quad, oct)-tree based grid (\S\ref{ssec:tree_structure}), adaptive mesh refinement (\S\ref{sssec:mesh_refinement}), and parallel task-based execution featuring hybrid parallelism through {\tt MPI} and {\tt OpenMP} (\S\ref{ssec:task_list}) lead \Athena to possess excellent scaling properties \cite{Stone:2020}. Here we confirm that, based on strong and weak scaling tests\footnote{In all the scaling tests presented $N_B = 16$, and we evolve for $20$ time-steps.}, extensions introduced in \GRAthena{} continue to preserve this. Our tests are performed in the context of evolution, with and without magnetic fields, of: BNS inspiral through to merger, and single NS (long term). Robust performance for demanding production grids, attained over multiple machines that involve a variety of architectures, is crucial for opening up the feasibility of very high resolution studies of these and related problems.

Scaling tests are performed on three clusters with different architectures: HLRS-HAWK\footnote{2 AMD EPYC 7742 CPUs / node, totalling 128 cores / node.};
SUPERMUC-NG\footnote{2 Intel Skylake Xeon Platinum 8174 CPUs / node, totalling, 48 cores / node.}; and Frontera\footnote{2 Intel 8280 Cascade Lake CPUs / node, totalling 56 cores / node.}. The physical and grid configurations employed here match those described for the relevant problem in prior sections.

Strong scaling tests are performed by fixing the {\tt Mesh} sampling $N_M$ of the unrefined grid and increasing the number of cores that the problem is solved on. For each choice of $N_M$ the physical extent of the innermost refined region is tuned to achieve a similar range of \MeshBlock /core
ratios across all series of resolutions.

The strong scaling efficiency is defined as $\eta_S := 1 - \frac{t - t_{\mathrm{ideal}}}{t_{\mathrm{ideal}}}$ where $t$ is the elapsed wallclock time and $t_\mathrm{ideal}$ is the time expected for perfect strong scaling, halving exactly when resources are doubled.

Strong scaling tests are performed on HAWK and SuperMUC, where a given node is saturated with $32$ {\tt MPI} tasks and $4$ {\tt OpenMP} threads / task, and $12$ {\tt MPI} tasks and $4$ {\tt OpenMP} threads / task respectively.

\begin{figure}[t]
  \sidecaption
  \centering
  \includegraphics[width=0.6\textwidth]{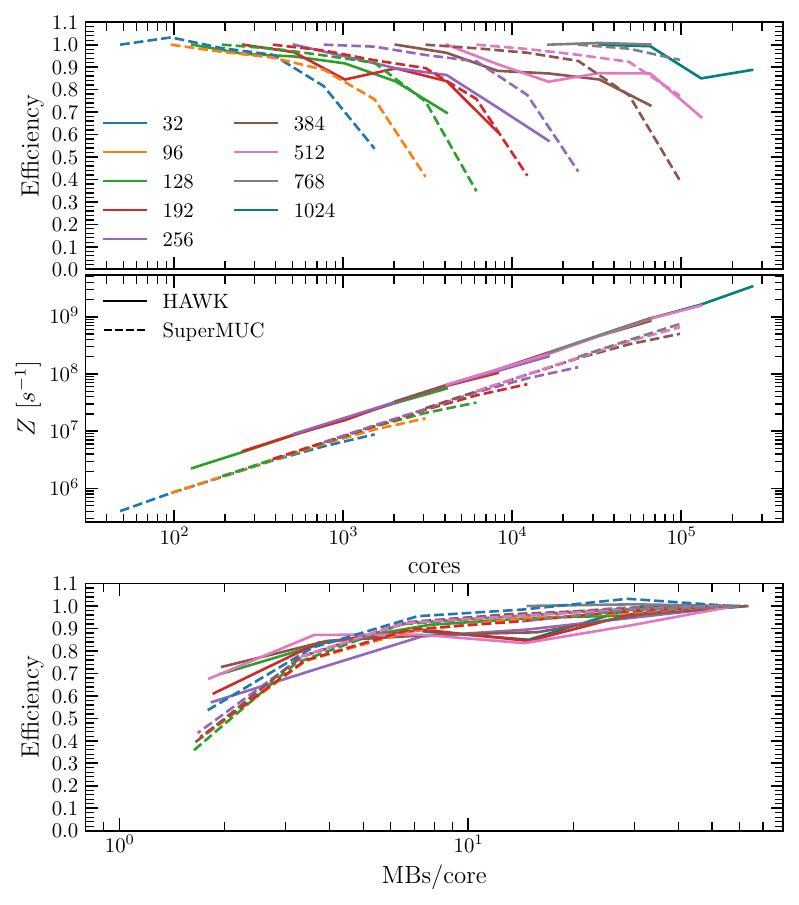}
  \caption{Strong scaling efficiency $\eta_S$ for the binary neutron star configuration with MHD. Results for HAWK and SuperMUC are included, up to $\mathcal{O}(10^5)$ cores. %
  Top: efficiency versus number of cores. Middle: zone cycles per second ($Z$) versus number of cores. %
  Bottom: efficiency in terms of the \MeshBlock{} to core ratio. %
  (Figure adapted from \cite{Cook:2023bag}).
   }
 \label{fig:strong_scaling}
\end{figure}
\begin{figure}[t]
  \sidecaption
  \centering
  \includegraphics[width=0.58\textwidth]{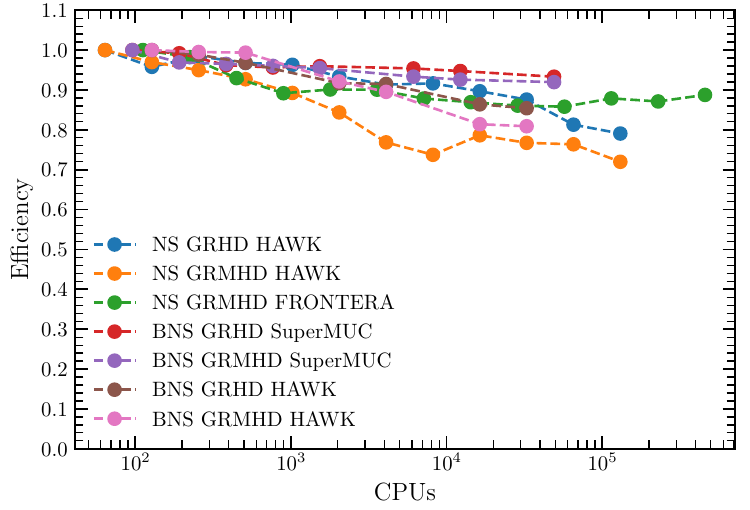}
  \caption{
  Weak scaling $\eta_W$ for SMR grids, over different target machines, for NS and BNS configurations with (GRMHD), and without (GRHD) magnetic fields.     (Figure adapted from \cite{Cook:2023bag}).
   }
 \label{fig:weak_scaling}
\end{figure}
In Fig.~\ref{fig:strong_scaling} we show the strong scaling performance for BNS with GRMHD. In the top panel, demonstrating efficiency against number of cores, we find efficiencies in excess of $80\%$ up to $\sim 10^5$ cores (gray and teal lines) for both SuperMUC (dashed lines) and HAWK (solid lines). An important aspect of maintaining high efficiencies is sufficiently saturating computational load \cite{Daszuta:2021ecf}.
This can be seen in Fig.~\ref{fig:strong_scaling} (lower) where the efficiency
on SuperMUC is $\gtrsim 90\%$ as long as $\mathrm{\MeshBlock{} /core} \gtrsim 6$, and the drop in efficiency on HAWK corresponding to this ratio dropping below $4$. The differences between machines are attributed to varied machine architectures. This causes a difference in the raw performance in terms of zone cycles per second ($Z$) (middle panel), where we observe runs on HAWK executing $\gtrsim 2$ faster than on SuperMUC. Comparable results found for simulations with magnetic fields disabled, and single NS configurations, have been omitted for brevity.

For weak scaling the executed zone cycles per second, $Z$, are measured. The expectation is that for perfect scaling, $Z$ should double every time the computational load and resources are concurrently doubled. The weak efficiency measure $\eta_W := 1 - \frac{Z - Z_{\mathrm{ideal}}}{Z_{\mathrm{ideal}}}$ captures this.

In Fig~\ref{fig:weak_scaling} we demonstrate this weak scaling performance on the full range of machines and problems discussed above. Observe that
scaling is maintained up to $\sim{}5\times 10^5$ CPU cores for the single star test at 89\% efficiency on 
Frontera, with efficiencies dropping to, at worst 72\% on HAWK. For binary tests we see an efficiency $\gtrsim 90\%$ up to $\sim{}5\times 10^4$ cores on SuperMUC-NG, with slightly lower efficiencies seen on HAWK for the same problem. This is also consistent with our previous vacuum sector tests \cite{Daszuta:2021ecf} and those for stationary space-times \cite{Stone:2020}.

\section{Summary and outlook}
Our overview of \GRAthena{} for GRMHD simulations of astrophysical flows on dynamical space-times described several novel aspects we have introduced with respect to the original \Athena{}. In particular our treatment of: \z4c{} coupling to GRMHD; equation of state; discretizations over differing grids including reconstruction; the constrained transport algorithm; conservative-to-primitive variable inversion; extraction of gravitational waves utilizing geodesic grids. When operation of these features is interwoven during GRMHD simulation, robust performance characteristics on multiple machines was demonstrated. Evolution on production grids in the absence of symmetries shows efficient use of exascale HPC architecture where: strong scaling efficiency in excess of $80\%$ up to $10^5$ CPU cores, and weak scaling efficiency of $89\%$ for $\sim{}5\times 10^5$ CPU cores is demonstrated.

Representative of the modeling challenges posed by astrophysical applications, benchmark problems  were utilized to establish simulation quality of \GRAthena{} with focus on the absence of any simplifying symmetry reductions. This involved evolution of: isolated equilibrium NS over the long-term; unstable, rotating NS through gravitational collapse and black hole remnant; BNS inspiral through merger. Gravitational waveforms are shown to be convergent as resolution is increased, and consistent in cross-code comparison with {\tt BAM}.

Anticipating full-scale simulations that seek to start tackling questions surrounding the open problems outlined in the introduction we presented novel simulations involving magnetic field instabilities. Long-term evolution of isolated magnetised NS with initially poloidal fields up to 68.7 ms extend our previous simulations \cite{Sur:2021awe} to include a dynamical space-time. Similar results for the growth of a toroidal field component were found, saturating at approximately $10\%$ of the total magnetic field energy. In contrast, with the current GRMHD treatment of \GRAthena{}, we find superior conservation of the internal energy of the NS, and total relative violations of the magnetic field divergence-free condition of the order of machine round-off. Similarly, we show that the KHI may be efficiently resolved through our AMR infrastructure. Indeed even initial low resolutions yield an amplification of a factor $\sim{}8$ with a single, additional level of refinement whilst utilizing approximately half the number of {\tt MeshBlock} objects at the moment of merger when contrasted with a comparable static mesh refinement approach.

Active work on \GRAthena{} is underway so as to incorporate high-order schemes \cite{Radice:2013hxh,Bernuzzi:2016pie,Doulis:2022vkx} essential for further improving waveform convergence and quality. The neutrino transport scheme developed by \cite{Radice:2021jtw} is presently being ported together with an improved treatment of weak reactions and reaction rates. In future we also plan to couple the recently developed radiation solvers of \cite{Bhattacharyya:2022bzf,White:2023wxh}. Once these developments are mature, it is our intention to make \GRAthena{} publicly available.

\begin{acknowledgement}
BD acknowledges funding from the EU H2020 under ERC Starting Grant, no.~BinGraSp-714626, and from the EU Horizon under ERC Consolidator Grant, no. InspiReM-101043372.
\end{acknowledgement}

\end{document}